\documentclass[aps,prl,reprint,superscriptaddress]{revtex4-2}
\usepackage{amsmath,amssymb,bm,graphicx}
\usepackage[colorlinks=true,linkcolor=blue,citecolor=blue,urlcolor=blue]{hyperref}

\usepackage[dvipsnames]{xcolor}
\usepackage{physics}

\begin{document}

\title{
{Tunable Mediated Interactions Near Spontaneous Symmetry Breaking: \\
From
Yukawa to Coulomb and Dzyaloshinskii--Moriya Interactions}
}

\author{Hoshu Hiyane}
\affiliation{Institut f\"{u}r Theoretische Physik, Leibniz Universit\"{a}t Hannover, Appelstr. 2, 30167 Hannover, Germany}

\author{Hiroyuki Tajima}
\affiliation{Department of Physics, The University of Tokyo, Tokyo 113-0033, Japan}
\affiliation{RIKEN Nishina Center, Wako 351-0198, Japan}
\affiliation{Quark Nuclear Science Institute, The University of Tokyo, Tokyo 113-0033, Japan}

\date{\today}

\begin{abstract}
We propose a realization of tunable long-range mediated interactions generated by Nambu--Goldstone (NG) modes near spontaneous symmetry breaking.
An effective exchange interaction emerges between two distinguishable impurities proportional to the retarded susceptibility mediated by the bath particles.
The interaction is mediated by emergent NG modes associated with symmetry breaking in systems, namely the magnon mode in a repulsive two-component Fermi or Bose gas, or the goldstino mode in a Bose--Fermi mixture.
When the NG mode is gapped by explicit symmetry breaking, the interaction has a Yukawa-type short-ranged interaction, whereas the gapless limit gives a Coulomb law.
Above the NG mode threshold, the Dzyaloshinskii–Moriya-like cross product coupling emerges in a long-range form.
Our proposal paves a way towards the realization of a quantum simulator of spin models and polarons with controllable non-local interactions.
\end{abstract}

\maketitle

\paragraph{Introduction.---}
Mediated interactions are among the most direct manifestations of collective excitations.
In condensed matter physics, it is known that phonons, the Nambu--Goldstone (NG) mode~\cite{PhysRev.117.648,PhysRev.127.965} associated with the spontaneous symmetry breaking (SSB) of the spatial translation symmetry, induce the attractive interaction between electrons, leading to superconductivity as described by the Bardeen--Cooper--Schrieffer theory~\cite{PhysRev.108.1175}.
In nuclear physics, pions, which can also be regarded as the NG mode associated with SSB of the chiral symmetry~\cite{PhysRevLett.4.380}, play a crucial role in understanding nuclear forces.
In particular, the one-pion exchange process leads to the so-called Yukawa interaction~\cite{yukawa1935interaction}, in which the interaction range is given by the pion mass.

Recently, the mediated interactions have been investigated in ultracold atoms with tunable settings~\cite{bloch2012quantum}.
A cold atomic mixture with a large population imbalance can be used to test how minority particles, that behave as impurities, are affected by a surrounding medium bath within the polaron description~\cite{baroni2024quantum,PhysRevA.110.030101}.
While impurities immersed in superfluids feel the Yukawa-type mediated interaction due to the exchange of a Bogoliubov phonons, which is a density excitation of the condensate medium~\cite{PhysRevLett.85.2418,PhysRevA.61.053601,PhysRevA.71.033605,PhysRevB.93.205144,PhysRevLett.121.013401,naidon2018two,Gomez-Lozada2025Apr,Hiyane2024Aug,hiyane2025condensate}, those immersed in a Fermi sea experience the Ruderman--Kittel--Kasuya--Yosida--type interaction~\cite{RudermanKittel1954,Kasuya1956,Yosida1957} that has been
discussed theoretically~\cite{pethick2008bose,PhysRevA.72.023616,PhysRevA.79.013629,PhysRevResearch.1.033177,PhysRevLett.129.083401,PhysRevA.110.033304} and
observed experimentally~\cite{desalvo2019observation,PhysRevLett.124.163401,baroni2024mediated,w7rd-2qpv}.

Fl\"{o}hrich, in his seminal work~\cite{frohlich1954electrons}, described two interacting electrons in a simple (longitudinal optical) phononic bath via a Coulomb law through an exchange of the bath phonons~\cite{frohlich1954electrons,PhysRev.99.1140,devreese2009frohlich,Devreese2010Dec}. 
Despite its simplification,
the effect of such long-range mediated interaction indeed emerges and is often large enough to modify the effective strength of the interaction among electrons~\cite{devreese2009frohlich,Devreese2010Dec,Filip2021Aug,Tubman2026Jun}.
However, tunable-range interactions mediated by the NG mode
have not yet been implemented in cold-atom setups.
Considering the controllability that cold atoms offer as a quantum simulator, 
the realization of tunable long-range mediated interaction is highly desirable.

\begin{figure}[t]
    \centering
    \includegraphics[width=\linewidth]{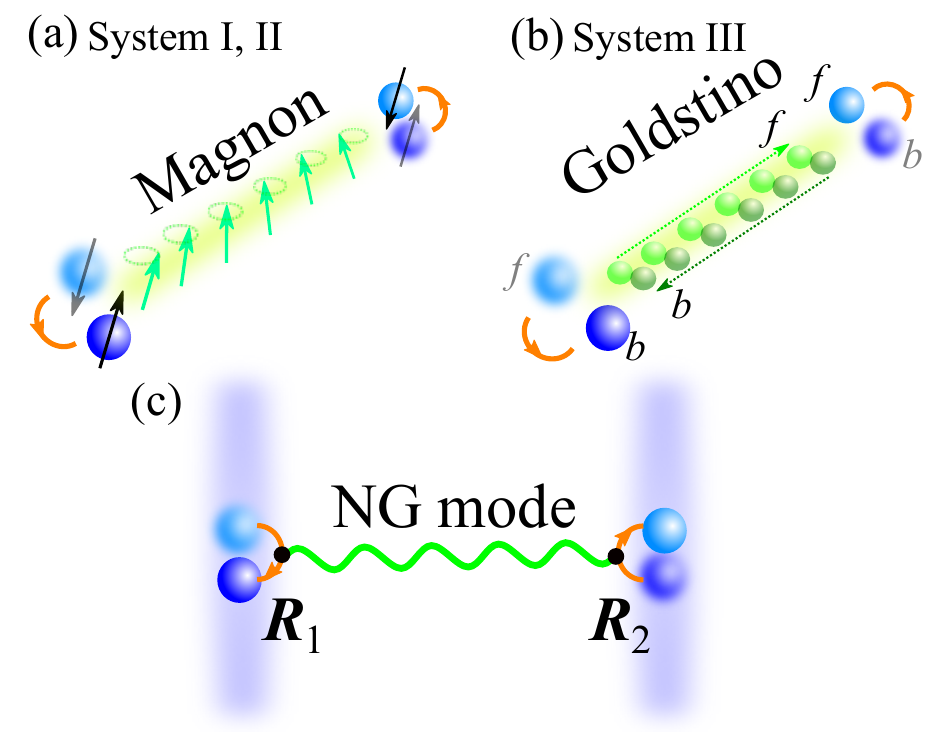}
    \caption{
    Schematics of mediated interactions between two impurities. 
    Three types of systems (I) Fermi--Fermi, (II) Bose--Bose, and (III) Bose--Fermi mixture are considered.
    (a) For a ferromagnetic medium (System I, II), a magnon excitation mediates 
    the spin-exchange interaction between two impurities.
    (b) For a Bose--Fermi mixture (System III), a goldstino excitation 
    mediates an interaction that exchanges the two statistically different impurities.
    (c) Proposed detection of long-range interactions mediated by the NG mode. 
    Two impurities are localized at positions $\bm{R}_1$ and $\bm{R}_2$ by the optical tweezers (represented by shaded area) and involve the state-exchange interactions with the medium, denoted by black dots.
    For Systems I or II, such localized impurities can be regarded as a quantum simulator for a spin system with tunable mediated interactions.
    }
    \label{fig:1}
\end{figure}

Here, we investigate two impurities immersed in a bath of particles that facilitate a long-range interaction mediated by the characteristic NG mode of the bath systems.
Our theory is based on an exchange interaction between bath particles and impurities generally; thus, we illustrate with three different kinds of systems, all of which possess type-B NG modes~\cite{PhysRevLett.108.251602,PhysRevLett.110.091601}.
(I) Two fermionic impurities with different internal states 
immersed in a bath of repulsive Fermi--Fermi mixtures.
The bath hosts a magnon as an elementary excitation due to the SSB of spin rotation~\cite{PhysRevLett.95.230403,Sandri2011,tajima2021non,PhysRevB.108.155303,
PhysRevLett.118.083602,valtolina2017exploring,PhysRevA.101.013603,PhysRevLett.129.203402}.
We show that the two fermionic impurities exchange the magnons, yielding an effective interaction with range being tunable from Yukawa to Coulomb law (see Fig.~\ref{fig:1}(a)).
(II) Similarly, immersing two bosonic impurities into the Bose--Bose mixture, where the spin-density mode can be regarded as a magnon-like excitation~\cite{recati2022coherently}, also facilitates the same kind of tunable effective interaction between impurities.
(III) Finally, we show that the tunable long-range interaction also emerges in Bose--Fermi mixtures, where the broken-supersymmetry yields a fermionic collective excitation called goldstino~\cite{PhysRevLett.100.090404,PhysRevA.92.063629,PhysRevA.93.033642,PhysRevA.96.063617,TajimaHidakaSatow2021,gazzillo2026inverse} as an NG mode.
Then, two bosonic and fermionic impurities immersed in such a bath can exchange goldstino, resulting in a mediated range-tunable interaction (see Fig.~\ref{fig:1}(b)).

Importantly, the range of the effective interaction is related to the energy difference between two internal states and, if exist, the energy gap of the NG mode due to the explicit SSB.
Applying our framework to localized spin systems, we show that a spin-flip interaction emerge with a tunable range from short-ranged Yukawa to long-ranged Coulomb-type.
Furthermore, the Dzyaloshinskii--Moriya (DM)-type coupling~\cite{dzyaloshinsky1958thermodynamic,Moriya1960} appears when the on-shell excitation of the NG mode is allowed.
While spin systems with long-range interactions 
have attracted interest recently~\cite{RevModPhys.93.025001,PhysRevLett.109.025303,Chen2025Jun,hsbt-c46n,75k9-hx9k}, our work provides a new realization route of
a quantum simulator for 
long-range interacting spin systems.
This simulator in turn provides a 
simple setup to detect such a tunable 
range interaction
via the time evolution of internal states of two pinned impurities (see also Fig.~\ref{fig:1}(c)).
The interaction-induced Rabi cycle distinguishes
either the Yukawa or Coulomb laws,
while the DM-interaction effect can be seen via the Ramsey interferometry.

\paragraph{Model for two impurities in a medium bath.---}
In the following, we use the convention of $\hbar=k_{\rm B}=1$, and the system volume is taken to be unity.
We consider two distinguishable impurities immersed in a medium bath near SSB described by the Hamiltonian
    $\hat{H}=\hat{H}_{\rm I}+\hat{H}_{\rm M}+\hat{H}_{\rm int}$,
where 
\begin{align}
\label{eq:h_I}
    \hat{H}_{\rm I}=\sum_{\sigma=\sigma_1,\sigma_2}
    \int d\bm{r} \,
    \hat{\psi}_{{\rm I},\sigma}^\dag(\bm{r})\left[-\frac{\nabla^2}{2m_\sigma}+U_{\sigma}(\bm{r})\right]\hat{\psi}_{{\rm I},\sigma}(\bm{r})
\end{align}
is the impurity Hamiltonian with a field $\psi_{{\rm I},\sigma}(\bm{r})$ of distinguishuable impurities with mass $m_\sigma$.
They are labeled by $\sigma_{1,2}=\uparrow,\downarrow$ for systems I and II, while $\sigma_{1,2}=b,f$ for system III indicating bosonic and fermionic fields.
We summarize the role of each operator in Tab.~\ref{tab:1}.
 $U_{\sigma}(\bm{r})$ corresponds to the single-particle trapping potential.
 $\hat{H}_{\rm M}$ 
 is the Hamiltonian of the medium bath consisting of field operators $\hat{\psi}_{{\rm M},\sigma}(\bm{r})$.
While the explicit form of $\hat H_{\rm M}$ is not necessary in deriving effective interaction among impurities, we show their details in the
supplemental material for completeness~\cite{suppmat}.
\begin{table}[bt]
    \centering
     \begin{ruledtabular}
    \begin{tabular}{c|c|c|c}
    $\bm{\psi}$ & (I) Fermi--Fermi & (II)  Bose--Bose & (III) Bose--Fermi \\
    \hline
        $\hat{\psi}_{{\rm I},\sigma_1}$ & $\hat{\psi}_{{\rm I},\uparrow}$ (Fermi) & $\hat{\psi}_{{\rm I},\uparrow}$ (Bose) & $\hat{\psi}_{{\rm I},b}$ (Bose)\\
        $\hat{\psi}_{{\rm I},\sigma_2}$ & $\hat{\psi}_{{\rm I},\downarrow}$ (Fermi)& $\hat{\psi}_{{\rm I},\downarrow}$ (Bose) & $\hat{\psi}_{{\rm I},f}$ (Fermi)\\
        $\hat{\psi}_{{\rm M},\sigma_1}$ & $\hat{\psi}_{{\rm M},\uparrow}$ (Fermi)& $\hat{\psi}_{{\rm M},\uparrow}$ (Bose) & $\hat{\psi}_{{\rm M},b}$ (Bose)  \\
        $\hat{\psi}_{{\rm M},\sigma_2}$ & $\hat{\psi}_{{\rm M},\downarrow}$ (Fermi)&$\hat{\psi}_{{\rm M},\downarrow}$ (Bose) &$\hat{\psi}_{{\rm M},f}$ (Fermi)\\
    \end{tabular}
     \end{ruledtabular}
    \caption{Correspondence between the general four-field notation and practical realizations in Fermi--Fermi, Bose--Fermi, and Bose--Bose mixtures. The labels ${\rm I}$ and ${\rm M}$ denote impurity and bath medium fields,
respectively.
The $\sigma_1$ and $\sigma_2$ 
are fixed to represent $\uparrow$ and $\downarrow$ for pseudospins (i.e., hyperfine states) in (I) Fermi--Fermi and (II) Bose--Bose mixtures, while they are $f$ and $b$ for (III) Bose--Fermi mixture, indicating the Fermi and Bose statistics of each particle.}
    \label{tab:1}
\end{table}

The exchange interaction $\hat{H}_{\rm int}$ between an impurity and a medium atoms in 
$\hat{H}$
plays an important role and is given by
\begin{align}
\label{eq:V}
    \hat{H}_{\rm int}=g\int d^3\bm{r} \, 
    [\hat{S}_{{\rm I},+}(\bm{r})
    \hat{S}_{{\rm M},-}(\bm{r})
  +\hat{S}_{{\rm M},+}(\bm{r})\hat{S}_{{\rm I},-}(\bm{r})],
\end{align}
where $g$ is a coupling constant and 
    $\hat{S}_{{\rm I},+}(\bm{r})=
    \hat{\psi}_{{\rm I},\sigma_1}^\dag(\bm{r})
    \hat{\psi}_{{\rm I},\sigma_2}(\bm{r})$ and 
    $\hat{S}_{{\rm M},-}(\bm{r})=
    \hat{\psi}_{{\rm M},\sigma_2}^\dag(\bm{r})
    \hat{\psi}_{{\rm M},\sigma_1}(\bm{r})$ are the ladder-type operators of impurities and medium atoms, with relations $\hat{S}_{{\rm I},-}(\bm{r})=\hat{S}_{{\rm I},+}^\dag(\bm{r})$ and $\hat{S}_{{\rm M},+}(\bm{r})=\hat{S}_{{\rm M},-}^\dag(\bm{r})$.
While each particle can interact via 
conventional density-density coupling, it does not play a major role in this proposal, as our aim is to study 
a mediated long-range force in the exchange process. 
$\hat{H}_{\rm int}$ changes the internal states of both impurity and medium atoms.
We note that such an interaction potential has already been discussed in detail 
in quantum mixtures~\cite{scazza2014observation,PhysRevA.93.043601,PhysRevB.97.155156,PhysRevA.98.023601,PhysRevLett.121.130403}.

\paragraph{Long-range mediated interaction between two impurities.---}
We derive the mediated interaction $\hat{V}_{\rm eff}$ between two impurities by tracing out the medium excitation~\cite{Tajima2025,v1y3-2b6r}.
To this end, let us consider
the partition function $Z$ of the total system: $Z=\int \mathcal{D}\bm{\psi}_{\rm I}\mathcal{D}\bm{\psi}_{\rm M} e^{-S}$, where $S=S_I+S_M+S_{\rm int}$ is the action corresponding to 
$\hat{H}$
and $\bm{\psi}_{\rm I(M)}$ collectively denotes all the fields of impurity (medium) particles (see Tab.~\ref{tab:1}).
The effective action of impurities $S_{\rm eff}$ is realized by integrating out the bath particles: $Z=\int \mathcal{D}\bm{\psi}_Ie^{-S_{\rm eff}[\bm{\psi}_I]}$, where
$e^{-S_{\rm eff}}=\int\mathcal{D}\bm{\psi}_Me^{-S_M}e^{-S_{\rm int}} =\ev{e^{-S_{\rm int}}}_{\rm M}$.
This expectation value can be perturbatively expanded by means of the cumulative expansion, and in its leading order, we find
$\ev{e^{-S_{\rm int}}}_{\rm M}=\exp[-\ev{S_{\rm int}}_{\rm M}+\ev{S^2_{\rm int}}_{\rm M}/2]$.
Noticing that $\langle {S}_{\rm int}\rangle_{\rm M}$ 
gives the renormalization of the single-impurity energy and is irrelevant for the two-impurity interaction,
we identify the effective action that describes mediated exchange interaction as
$S_{\rm eff}=- \ev{S^2_{\rm int}}_{\rm M}/2$. This yields the effective interaction among impurities,
\begin{align}
\label{eq:V_eff}
    \hat{V}_{\rm eff}=-\frac{1}{2\beta}\int_0^{\beta}d\tau_1\int_0^{\beta}d\tau_2
    \langle
    T_\tau
    [\hat{H}_{\rm int}(\tau_1)\hat{H}_{\rm int}(\tau_2)
    ]
    \rangle_{\rm M},
\end{align}
with the inverse temperature $\beta=1/T$, where $T_\tau$ is the imaginary-time ordering product.
The expectation value in the right-hand side
can be evaluated by using the standard thermal Green's function technique~\cite{FetterWalecka1971}.
Substituting Eq.~\eqref{eq:V} into Eq.~\eqref{eq:V_eff} and using the spatial and temporal translational invariance of the medium,
we obtain
\begin{align}
\label{eq:Def_effective_interaction}
    \hat{V}_{\rm eff}
    &=-\frac{|g|^2}{2\beta}\int d^3\bm{r}_1
    \int d^3\bm{r}_2\,
    \int_0^{\beta} d\tau_1
    \int_0^{\beta} d\tau_2
    \,
    \nonumber\\
    &\quad
     \chi(\bm{r}_1-\bm{r}_2,\tau_1-\tau_2)
    \hat{S}_{{\rm I,-}}(\bm{r}_1,\tau_1)
    \hat{S}_{{\rm I,+}}(\bm{r}_2,\tau_2)
    +{\rm h.c.,}
\end{align}
where $\hat{S}_{{\rm I},\pm}(\bm{r},\tau)=e^{\tau \hat{H}_{\rm I}}\hat{S}_{{\rm I},\pm}(\bm{r})e^{-\tau\hat{H}_{\rm I}}$ and
\begin{align}
    \chi(\bm{r},\tau)=-\langle T_\tau[\hat{S}_{{\rm M,}+}(\bm{r},\tau)\hat{S}_{{\rm M},-}(\bm{0},0)]\rangle_{\rm M}
\end{align}
is the imaginary-time susceptibility of the medium bath.

\paragraph{Long-range interaction between two localized impurities.---}
The effective interaction~\eqref{eq:Def_effective_interaction} couples two impurities via a Yukawa-type short-range, Coulomb-type long-range, and oscillatory long-range interaction.
In the following, we illustrate this by two impurities loaded in an optical tweezers schematically illustrated in Fig.~\ref{fig:1}(c).
The tweezers localize impurities at $\bm{r}=\bm{R}_1$ and $\bm{r}=\bm{R}_2$ with different internal states, and the Hamiltonian that describes impurities is
\begin{align}
\label{eq:HeffForLocalizedImp}
    \hat{H}_{\rm eff}=& \hat H_{\rm I}+\hat V_{\rm eff},
    \quad \hat{H}_{\rm I}\simeq 
    \sum_{j=1,2}\Delta_{j}\hat{s}_j^{+}\hat{s}_{j}^{-},
\end{align}
where $\Delta_j$ is the energy difference between two states $|\sigma_1\rangle_{\bm{R}_j}$ and $|\sigma_2\rangle_{\bm{R}_j}$ of the impurity at position $\bm R_j$~\footnote{We note that $\Delta_j$ is the level splitting of the full localized single impurity Hamiltonian, including the bare energy difference between the two states, mean-field energy shifts from diagonal density-density coupling, and the one-body energy shift arising from $\ev*{\hat H_{\rm int}}_M$ induced by the exchange interaction.
}.
The operators $\hat{s}_{j}^{\pm}$ flips states as
$\hat{s}_{j}^{+}|\sigma_1\rangle_{\bm{R}_j}=|\sigma_2\rangle_{\bm{R}_j}$ and $\hat{s}_{j}^{-}|\sigma_2\rangle_{\bm{R}_j}=|\sigma_1\rangle_{\bm{R}_j}$.
In this case, we obtain 
$\hat{S}_{{\rm I},\pm}(\bm{r},\tau)\simeq
\sum_{j=1,2}\delta(\bm{r}-\bm{R}_j)e^{\pm\Delta_j\tau}\hat{s}_j^{\pm}$ where we use $e^{\hat{H}_{\rm I}\tau}\hat{s}_j^{\pm}e^{-\hat{H}_{\rm I}\tau}=e^{\pm\Delta_j\tau}\hat{s}_j^{\pm}$.
Then, the effective interaction is given by~\footnote{In the second line of Eq.~\eqref{eq:j},
we use the analytic continuation from $\chi(\bm{R},i\omega_n)=\int_0^{\beta}d\tau\,e^{-i\omega_n\tau}\chi(\bm{R},\tau)$ with the Matsubara frequency $\omega_n$ to the retarded susceptibility $\chi^{\rm ret}(\bm{R},\omega)\equiv \chi(\bm{R},i\omega_n\rightarrow\omega+i0^+)$ in the frequency representation.}
\begin{align}
    \hat{V}_{\rm eff}&\simeq J(\bm{R})\hat{s}_{1}^{-}\hat{s}_{2}^{+}+{\rm h.c.},\label{eq:10}\\
    J(\bm{R})&=-|g|^2\int_0^{\beta}d\tau\,e^{-(\Delta_1-\Delta_2)\tau}\chi(\bm{R},\tau)\nonumber \\
    &=-|g|^2\chi^{\rm ret}(\bm{R},\omega=\Delta_1-\Delta_2)\label{eq:j},
\end{align}
with $J$ being a strength of the
coupling as a function of the relative distance $\bm{R}=\bm{R}_1-\bm{R}_2$.
We emphasize that Eq.~\eqref{eq:j} is
valid for both bosonic and fermionic collective excitations tabulated in Tab.~\ref{tab:1}.
Depending on the statistics of atoms, the retarded susceptibility in the time, 
$\chi^{\rm ret}(\bm{R},t)=-i\theta(t)\langle[\hat{S}_{{\rm M},+}(\bm{R},t),\hat{S}_{{\rm M},-}(\bm{0},0)]_{\eta}\rangle_{\rm M}$ requires the 
fermionic/bosonic
commutation relation $[\hat{A},\hat{B}]_\pm
=\hat{A}\hat{B}\pm\hat{B}\hat{A}$ 
for 
fermionic/bosonic
operators $\hat{A}$ and $\hat{B}$.

\begin{figure}[t]
    \centering
    \includegraphics[width=\linewidth]{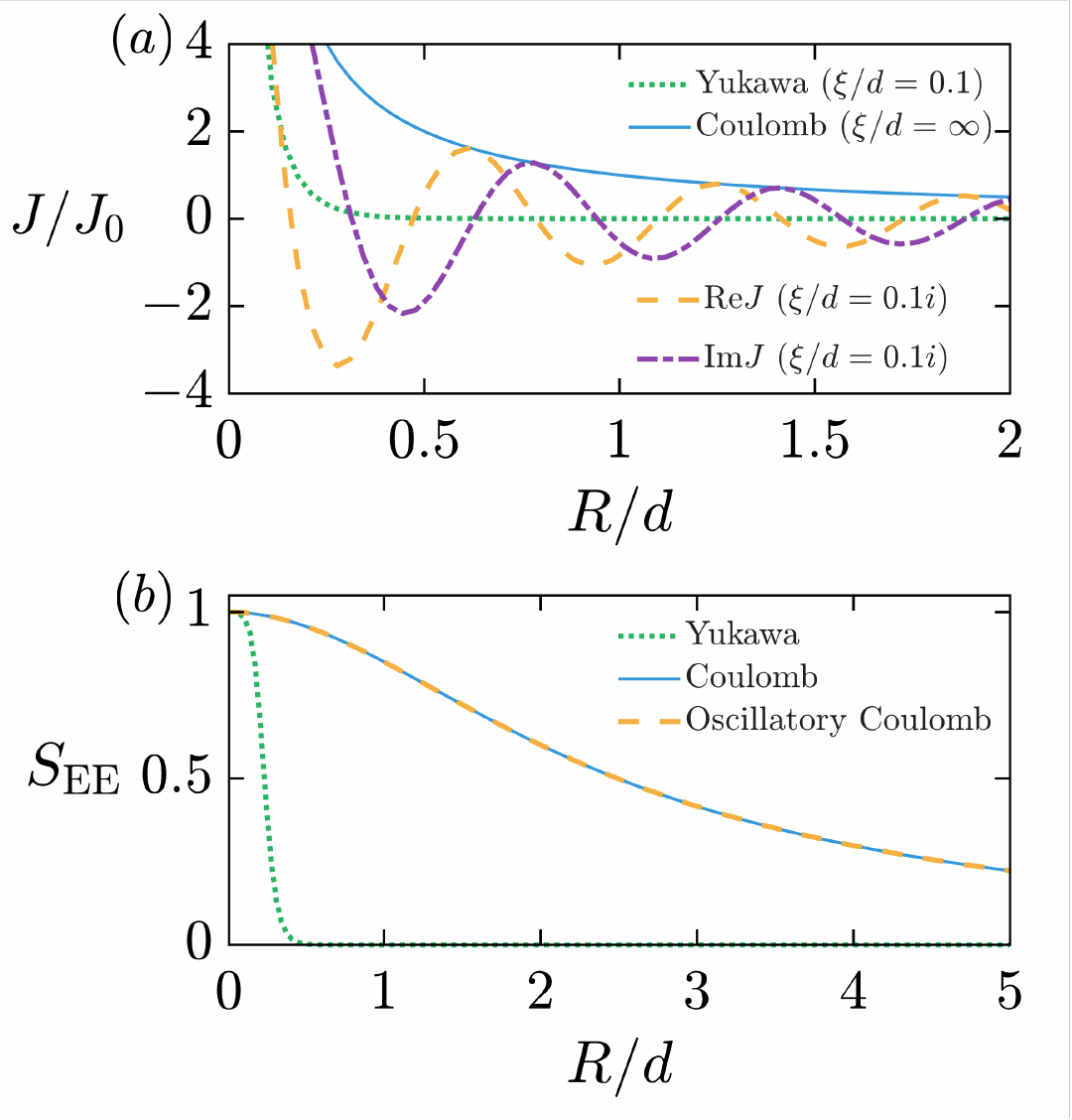}
    \caption{
    (a) Long-range mediated interaction $J(\bm{R})$ between impurities.
    The range of interaction $\xi=\sqrt{D/(\omega_{\rm gap}-\Delta_1+\Delta_2)}$ can be controlled by tuning gap size $\omega_{\rm gap}$, and energy difference between two states $\Delta_{1,2}$ of two impurities.
    The Coulomb ($\xi/d=\infty$), the Yukawa ($\xi/d=0.1$), and the oscillatory Coulomb case ($\xi=0.1i$) are visualized.
    In particular, ${\rm Re}\, J$ and ${\rm Im}\, J$ correspond to the XX and DM-type coupling, respectively.
    (b) Entanglement entropy $S_{\rm EE}$ of two localized impurities in a ground state of the effective Hamiltonian~\eqref{eq:HeffForLocalizedImp} with $\Delta_1=-\Delta_2=1/2$ in units of $J_0$.
    }
    \label{fig:2}
\end{figure}

Importantly, $\chi^{\rm ret}(\bm{R},\omega)$ is directly associated with the NG mode.
When the medium bath exhibits the type-B NG mode,
the Fourier component $\chi^{\rm ret}(\bm{q},\omega)$ can be written as~\cite{PhysRevD.91.056006}
\begin{align}
\label{eq:chi_ret}
    \chi^{\rm ret}(\bm{q},\omega)\simeq\frac{Z}{\omega-\omega_{\rm gap}-Dq^2+i\Gamma(\bm{q},\omega)},
\end{align}
where $Z$, $\omega_{\rm gap}$, $D$, and $\Gamma(\bm{q},\omega)$ are the pole residue, the excitation gap due to the explicit symmetry breaking, the dispersion coefficient, and the damping rate, respectively.
These parameters are directly associated with the medium properties.
For instance,
the dispersion parameters are given 
in  Refs.~\cite{PhysRev.170.576,Sandri2011,tajima2021non,PhysRevB.108.155303,suppmat} for the magnon in a two-component Fermi gas and in Refs.~\cite{PhysRevLett.100.090404,PhysRevA.92.063629,PhysRevA.93.033642,PhysRevA.96.063617,TajimaHidakaSatow2021,gazzillo2026inverse,suppmat} for the goldstino in a Bose-Fermi mixture.
Assuming the long lifetime of NG modes with weak damping (i.e., $\Gamma(\bm{0},0)\simeq 0$ in the low-energy region away from the continuum~\footnote{{
This condition is not strict and can be achieved, e.g.\ in ${}^{173}$Yb-${}^{174}$Yb or ${}^{40}$K-${}^{41}$K for the goldstino mode in the Bose--Fermi mixture~\cite{TajimaHidakaSatow2021} 
and for magnons in a Fermi--Fermi mixture~\cite{Sandri2011}, with inter-component repulsions.}}),
we obtain
\begin{align}
\label{eq:ExchangeIntStrength_JR}
    J(\bm{R})\simeq \frac{|g|^2Z}{4\pi D}\frac{e^{-R/\xi}}{R},
    \quad
    \xi=\sqrt{\frac{D}{\omega_{\rm gap}-\Delta_1+\Delta_2}}.
\end{align}

Figure~\ref{fig:2}(a) shows $J(\bm{R})$ as a function of $R/d\equiv|\bm{R}|/d$, where $d=\rho^{-1/3}$ is the mean interparticle distance of the medium bath with a number density $\rho$.
Depending on the system parameters, three types of interactions emerge;
For the case of $\omega_{\rm gap}-\Delta_1+\Delta_2>0$,
$\xi$ is a real value characterizing the range of the interaction and thus $J(\bm{R})$ is in the form of the Yukawa interaction $J_0e^{-R/\xi}/R$, where $J_0=|g|^2Z/4\pi D$.
However, at $\omega_{\rm gap}-\Delta_1+\Delta_2=0$,
the interaction range $\xi$ reaches to infinity and then, $J(\bm{R})$ exhibits the Coulomb-type power law $1/R$.
In this way, the long-range mediated interaction $J(\bm{R})$ with engineered length $\xi$ can be realized in the vicinity of SSB.
By preparing the localized impurity array with optical tweezers, one can realize a spin systems with long-range interaction described by Hamiltonian Eq.~\eqref{eq:HeffForLocalizedImp}-\eqref{eq:10} with the  strength of interaction Eq.~\eqref{eq:ExchangeIntStrength_JR}, where the SSB of the medium controls the interaction range.

Meanwhile, at $\omega_{\rm gap}-\Delta_1+\Delta_2<0$,
$\xi$ becomes a pure imaginary
and $J(\bm{R})$ exhibits a complex-valued 
$e^{iR/|\xi|}/R$
which we call oscillatory-Coulomb.
In contrast to the Yukawa case, such on-shell excitation of the NG mode is allowed at $\omega_{\rm gap}-\Delta_1+\Delta_2<0$.
This leads to the dissipation of the impurities' energy towards the medium excitation given by ${\rm Im}\, J(\bm{R})$ 
along with its inverse process as represented by the Hermiticity of $\hat{V}_{\rm eff}$ in Eq.~\eqref{eq:10}.
$|J(\bm{R})|$ is equal to that of the Coulomb-type case, and the oscillatory nature of $J(\bm{R})$ remains finite at large relative distance $R$.
The physical meaning of the oscillatory-Coulomb interaction becomes clearer by defining $\hat{s}_i^{x}=(\hat{s}_i^{+}+\hat{s}_i^{-})/2$ and
$\hat{s}_i^{y}=(\hat{s}_i^{+}-\hat{s}_i^{-})/2i$.
With these,
Eq.~\eqref{eq:10} can be rewritten as
\begin{align}
    \hat{V}_{\rm eff}
    \simeq
    2{\rm Re}\,J(\bm{R})
    (\hat{s}_1^x\hat{s}_2^x+\hat{s}_1^y\hat{s}_2^y)
    -
    2{\rm Im}\,J(\bm{R})
    (\hat{\bm s}_1\times\hat{\bm s}_2)_z.
\end{align}
It should be stressed 
that ${\rm Re}\,J(R)$ defines the strength of XX-type interaction, whereas ${\rm Im}\,J(\bm{R})$ gives the DM-type coupling strength, 
both of which are long-ranged as shown in Fig.~\ref{fig:2}(a).
Therefore, utilising system I or II, one can realize a quantum simulator for spin systems with a tunable range from Yukawa to Coulomb and XX/DM interactions.

To visualize the effect of long-range interaction between two impurities, we 
diagonalize the effective Hamiltonian~\eqref{eq:HeffForLocalizedImp}, \eqref{eq:10}, and \eqref{eq:ExchangeIntStrength_JR}, and obtain the two-body ground state $\ket{\Psi(\bm R_1,\bm R_2)}$.
Figure~\ref{fig:2}(b) shows the entanglement entropy $S_{\rm EE}=-\Tr \rho(\bm{R}_1)\log_2\rho(\bm{R}_1)$, where $\rho=\Tr_2\ket{\Psi}\bra{\Psi}$ is the one-body density matrix of impurities obtained by tracing out the particle at the position $\bm R_2$.
For all cases, the two impurities are highly correlated within $R\ll d$, due to the bath-mediated interaction.

However, as two impurities become far apart, the impurities interacting with the Yukawa law quickly become nearly separable; The entanglement rapidly vanishes before reaching even the averaged interparticle distances $d$.
On the other hand, the impurities interacting via the Coulomb law are correlated much longer than $d$, and so is the oscillatory-Coulomb system.
In fact, these two are exactly on top of each other for the ground state of two localized impurities as considered here.
This can be seen by writing the off-diagonal coupling as $J(\bm R)=|J|e^{i\theta}$ and $J^*(\bm R)=|J|e^{-i\theta}$.
Then, introducing the unitary transformation $\hat U={\rm diag}(e^{i\theta},1)$,
the transformed Hamiltonian $\hat U^\dagger \hat H_{\rm eff}\hat U$ does not depends on $\theta$.
Therefore, even for the oscillatory Coulomb case, $S_{\rm EE}$ remains finite and is equivalent to that of the Coulomb case (i.e., $\xi/d=\infty$), indicating that the impurities are entangled at long distances regardless of the imaginary part of $J(\bm{R})$.
On the other hand, in many-impurity arrays, $\theta$ may remain finite and thus lead to genuine chiral spin dynamics related to the DM-type coupling.

\begin{figure}[t]
    \centering
    \includegraphics[width=\linewidth]{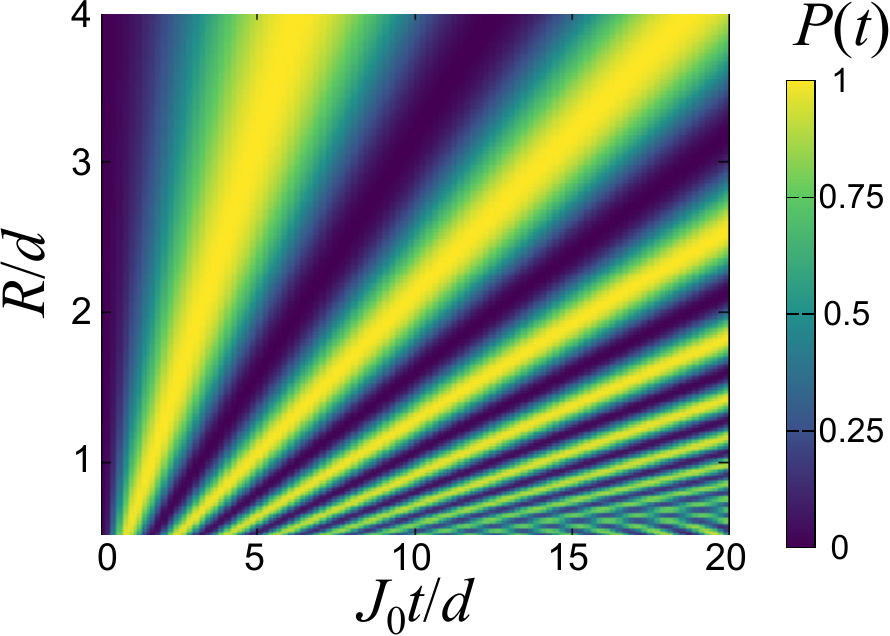}
    \caption{Transition probability $P(t)$ at different $R/d$ for $\xi=\infty$, where Coulomb-type long range exchange interaction emerge. 
    The short-distance region $R/d \leq 0.5$ is omitted because
$P(t)$ oscillates too rapidly to be resolved on this scale. 
}
    \label{fig:3}
\end{figure}

In the following, we propose a method 
to detect $J(\bm{R})$. Here, it is useful to consider the time evolution of the impurities.
Let $|a\rangle\equiv|\sigma_1\rangle_{\bm{R}_1}|\sigma_2\rangle_{\bm{R}_2}$, and 
 $|b\rangle\equiv |\sigma_2\rangle_{\bm{R}_1}|\sigma_1\rangle_{\bm{R}_2}$.
We initially prepare two impurities at $\ket{\Psi(0)}=\ket{a}$, and let it evolve by $\hat H_{\rm eff}$.
The mediated interaction generates a spin-flip dynamics with
$|\Psi(t)\rangle=c_a(t)|a\rangle+c_b(t)|b\rangle$, and 
the transition probability $P(t)\equiv |c_b(t)|^2$ 
shows interaction driven Rabi oscillation,
\begin{align}
    P(t)=\frac{|J(\bm{R})|^2}{|J(\bm{R})|^2+\omega_0^2/4}
    \sin^2\left[t\sqrt{|J(\bm{R})|^2+\omega_0^2/4}\right],
\end{align}
where $\omega_0=\Delta_1-\Delta_2$.
In particular, at $\omega_0=0$ (i.e., $\Delta_1=\Delta_2$),
one can obtain $|J(\bm{R})|$ from the oscillation period of $P(t)=\sin^2(|J(\bm{R})|t)$.
Figure~\ref{fig:3} shows $P(t)$ for the Coulomb case $J(\bm{R})=J_0/R$ at $\omega_0=0$.
While $P(t)= 1$ is achieved at $J_0 t/R=(n+1/2)\pi$ (where $n$ is an integer), $P(t)=0$ is found at $J_0 t/R=n\pi$.
For system III, this is a swap oscillation between bosons and fermions, in which the two statistically different particles at position $\bm R_{1}$ or $\bm R_2$ are exchanged in time ($|b\rangle_{\bm{R}_1}\ket{f}_{\bm{R}_2}\leftrightarrow|f\rangle_{\bm{R}_1}\ket{b}_{\bm{R}_2}$) in contrast to the spin-flop process in system I or II.
Such a conversion has not yet been realized and could be a crucial key in observing the goldstino excitation in cold atoms, as this occurs solely due to the goldstino-mediated interaction.

While $P(t)$ measures $|J(\bm R)|$, the phase
of $J(\bm R)$ can be accessed by Ramsey interferometry in the
two-state subspace, which has also been used for the hyperfine clock transition in the presence of the fermionic medium~\cite{PhysRevLett.124.163401}.
Preparing two impurities in a coherent superposition
of $\ket{\Psi_{\rm R}}=(|a\rangle+e^{i\Theta}|b\rangle)/\sqrt2$, and reading out $|\Psi(t)\rangle$ with a phase $\Theta$, one can measure the
coherence between $|a\rangle$ and $|b\rangle$ and hence $\theta\equiv {\rm Arg}\,J(\bm{R})$, as given by $\bra{\Psi_{\rm R}}e^{-i\hat{H}_{\rm eff}t}|a\rangle=\frac{1}{2}+\frac{1}{2}\sin(2|J(\bm{R})|t)\sin(\theta-\Theta)$.

\paragraph{Summary and outlook.---}
We have proposed a method to realize tunable long-range interactions mediated by the gapped and gapless NG modes.
Considering two localized impurities coupled with the NG mode in the medium bath,
we explicitly showed that the mediated interaction between the impurities exhibits Coulomb and Yukawa scaling laws with respect to the relative distance, depending on the excitation gap of the NG mode controlled by the explicit symmetry-breaking parameter.
Above the NG mode threshold, the long-range DM-type coupling appears together with the XX exchange coupling.

Our result opens a new route to realize tunable long-range spin systems by applying the present framework to arrays of localized impurities.
As a simplest demonstration, we have shown that the mediated exchange between two localized impurities
generates entanglement whose spatial profile reflects the tunable
interaction range.
The exchange process of the two localized impurities
during the time evolution allows one to probe the magnitude of mediated interaction, while the Ramsey fringe reads the phase.
For Bose--Fermi mixtures, this is a swap oscillation of bosonic and fermionic impurities, which may lead to the first observation of the goldstino excitation in the cold atoms.

On the other hand,
our work can also be applied to the system with optical lattices.
In the Mott insulating regime with magnons~\cite{PhysRevLett.103.110401,PhysRevLett.133.163402,PhysRevA.111.033312} and goldstinos~\cite{PhysRevB.109.085141,PhysRevB.111.075136}, the continuum effect could be safely neglected.
While we focused on emergent exchange interactions, the inherent density-density mediated interaction can also co-exist, and would play an important role in determining the physical properties of itinerant impurities.
A rich physics with new types of bound states and emergent phases is expected due to the interplay among different mediated interactions.
Furthermore, it would be interesting to extend our work to the three-body interaction by using the triad configuration~\cite{PhysRevLett.124.073401}.

Moreover, the goldstino-mediated exchange process considered here has a similarity to the Majorana-neutrino-exchange process considered in nuclear physics~\cite{Baym2025Mar},
from the perspective of the exchange process of internal states (e.g., isospin) mediated by fermionic excitation.
Engineering a quantum simulator with the fermion-mediated exchange process for the macroscopic neutrinoless double beta decay would be an intriguing future work.

\begin{acknowledgments}
This work was supported by 
Japan Society for the Promotion of Science (JSPS) Grants-in-Aid for Scientific Research (KAKENHI) Grant Nos.~JP22K13981, JP23K22429, and JP26K07063.
\end{acknowledgments}

\bibliographystyle{apsrev4-1}
\bibliography{Manuscript}

\begin{thebibliography}{81}%
\makeatletter
\providecommand \@ifxundefined [1]{%
 \@ifx{#1\undefined}
}%
\providecommand \@ifnum [1]{%
 \ifnum #1\expandafter \@firstoftwo
 \else \expandafter \@secondoftwo
 \fi
}%
\providecommand \@ifx [1]{%
 \ifx #1\expandafter \@firstoftwo
 \else \expandafter \@secondoftwo
 \fi
}%
\providecommand \natexlab [1]{#1}%
\providecommand \enquote  [1]{``#1''}%
\providecommand \bibnamefont  [1]{#1}%
\providecommand \bibfnamefont [1]{#1}%
\providecommand \citenamefont [1]{#1}%
\providecommand \href@noop [0]{\@secondoftwo}%
\providecommand \href [0]{\begingroup \@sanitize@url \@href}%
\providecommand \@href[1]{\@@startlink{#1}\@@href}%
\providecommand \@@href[1]{\endgroup#1\@@endlink}%
\providecommand \@sanitize@url [0]{\catcode `\\12\catcode `\$12\catcode `\&12\catcode `\#12\catcode `\^12\catcode `\_12\catcode `\%12\relax}%
\providecommand \@@startlink[1]{}%
\providecommand \@@endlink[0]{}%
\providecommand \url  [0]{\begingroup\@sanitize@url \@url }%
\providecommand \@url [1]{\endgroup\@href {#1}{\urlprefix }}%
\providecommand \urlprefix  [0]{URL }%
\providecommand \Eprint [0]{\href }%
\providecommand \doibase [0]{http://dx.doi.org/}%
\providecommand \selectlanguage [0]{\@gobble}%
\providecommand \bibinfo  [0]{\@secondoftwo}%
\providecommand \bibfield  [0]{\@secondoftwo}%
\providecommand \translation [1]{[#1]}%
\providecommand \BibitemOpen [0]{}%
\providecommand \bibitemStop [0]{}%
\providecommand \bibitemNoStop [0]{.\EOS\space}%
\providecommand \EOS [0]{\spacefactor3000\relax}%
\providecommand \BibitemShut  [1]{\csname bibitem#1\endcsname}%
\let\auto@bib@innerbib\@empty
\bibitem [{\citenamefont {Nambu}(1960{\natexlab{a}})}]{PhysRev.117.648}%
  \BibitemOpen
  \bibfield  {author} {\bibinfo {author} {\bibfnamefont {Y.}~\bibnamefont {Nambu}},\ }\href {\doibase 10.1103/PhysRev.117.648} {\bibfield  {journal} {\bibinfo  {journal} {Phys. Rev.}\ }\textbf {\bibinfo {volume} {117}},\ \bibinfo {pages} {648} (\bibinfo {year} {1960}{\natexlab{a}})}\BibitemShut {NoStop}%
\bibitem [{\citenamefont {Goldstone}\ \emph {et~al.}(1962)\citenamefont {Goldstone}, \citenamefont {Salam},\ and\ \citenamefont {Weinberg}}]{PhysRev.127.965}%
  \BibitemOpen
  \bibfield  {author} {\bibinfo {author} {\bibfnamefont {J.}~\bibnamefont {Goldstone}}, \bibinfo {author} {\bibfnamefont {A.}~\bibnamefont {Salam}}, \ and\ \bibinfo {author} {\bibfnamefont {S.}~\bibnamefont {Weinberg}},\ }\href {\doibase 10.1103/PhysRev.127.965} {\bibfield  {journal} {\bibinfo  {journal} {Phys. Rev.}\ }\textbf {\bibinfo {volume} {127}},\ \bibinfo {pages} {965} (\bibinfo {year} {1962})}\BibitemShut {NoStop}%
\bibitem [{\citenamefont {Bardeen}\ \emph {et~al.}(1957)\citenamefont {Bardeen}, \citenamefont {Cooper},\ and\ \citenamefont {Schrieffer}}]{PhysRev.108.1175}%
  \BibitemOpen
  \bibfield  {author} {\bibinfo {author} {\bibfnamefont {J.}~\bibnamefont {Bardeen}}, \bibinfo {author} {\bibfnamefont {L.~N.}\ \bibnamefont {Cooper}}, \ and\ \bibinfo {author} {\bibfnamefont {J.~R.}\ \bibnamefont {Schrieffer}},\ }\href {\doibase 10.1103/PhysRev.108.1175} {\bibfield  {journal} {\bibinfo  {journal} {Phys. Rev.}\ }\textbf {\bibinfo {volume} {108}},\ \bibinfo {pages} {1175} (\bibinfo {year} {1957})}\BibitemShut {NoStop}%
\bibitem [{\citenamefont {Nambu}(1960{\natexlab{b}})}]{PhysRevLett.4.380}%
  \BibitemOpen
  \bibfield  {author} {\bibinfo {author} {\bibfnamefont {Y.}~\bibnamefont {Nambu}},\ }\href {\doibase 10.1103/PhysRevLett.4.380} {\bibfield  {journal} {\bibinfo  {journal} {Phys. Rev. Lett.}\ }\textbf {\bibinfo {volume} {4}},\ \bibinfo {pages} {380} (\bibinfo {year} {1960}{\natexlab{b}})}\BibitemShut {NoStop}%
\bibitem [{\citenamefont {Yukawa}(1935)}]{yukawa1935interaction}%
  \BibitemOpen
  \bibfield  {author} {\bibinfo {author} {\bibfnamefont {H.}~\bibnamefont {Yukawa}},\ }\href@noop {} {\bibfield  {journal} {\bibinfo  {journal} {Proceedings of the Physico-Mathematical Society of Japan. 3rd Series}\ }\textbf {\bibinfo {volume} {17}},\ \bibinfo {pages} {48} (\bibinfo {year} {1935})}\BibitemShut {NoStop}%
\bibitem [{\citenamefont {Bloch}\ \emph {et~al.}(2012)\citenamefont {Bloch}, \citenamefont {Dalibard},\ and\ \citenamefont {Nascimbene}}]{bloch2012quantum}%
  \BibitemOpen
  \bibfield  {author} {\bibinfo {author} {\bibfnamefont {I.}~\bibnamefont {Bloch}}, \bibinfo {author} {\bibfnamefont {J.}~\bibnamefont {Dalibard}}, \ and\ \bibinfo {author} {\bibfnamefont {S.}~\bibnamefont {Nascimbene}},\ }\href@noop {} {\bibfield  {journal} {\bibinfo  {journal} {Nature Physics}\ }\textbf {\bibinfo {volume} {8}},\ \bibinfo {pages} {267} (\bibinfo {year} {2012})}\BibitemShut {NoStop}%
\bibitem [{\citenamefont {Baroni}\ \emph {et~al.}(2024{\natexlab{a}})\citenamefont {Baroni}, \citenamefont {Lamporesi},\ and\ \citenamefont {Zaccanti}}]{baroni2024quantum}%
  \BibitemOpen
  \bibfield  {author} {\bibinfo {author} {\bibfnamefont {C.}~\bibnamefont {Baroni}}, \bibinfo {author} {\bibfnamefont {G.}~\bibnamefont {Lamporesi}}, \ and\ \bibinfo {author} {\bibfnamefont {M.}~\bibnamefont {Zaccanti}},\ }\href@noop {} {\bibfield  {journal} {\bibinfo  {journal} {Nature Reviews Physics}\ }\textbf {\bibinfo {volume} {6}},\ \bibinfo {pages} {736} (\bibinfo {year} {2024}{\natexlab{a}})}\BibitemShut {NoStop}%
\bibitem [{\citenamefont {Paredes}\ \emph {et~al.}(2024)\citenamefont {Paredes}, \citenamefont {Bruun},\ and\ \citenamefont {Camacho-Guardian}}]{PhysRevA.110.030101}%
  \BibitemOpen
  \bibfield  {author} {\bibinfo {author} {\bibfnamefont {R.}~\bibnamefont {Paredes}}, \bibinfo {author} {\bibfnamefont {G.}~\bibnamefont {Bruun}}, \ and\ \bibinfo {author} {\bibfnamefont {A.}~\bibnamefont {Camacho-Guardian}},\ }\href {\doibase 10.1103/PhysRevA.110.030101} {\bibfield  {journal} {\bibinfo  {journal} {Phys. Rev. A}\ }\textbf {\bibinfo {volume} {110}},\ \bibinfo {pages} {030101} (\bibinfo {year} {2024})}\BibitemShut {NoStop}%
\bibitem [{\citenamefont {Heiselberg}\ \emph {et~al.}(2000)\citenamefont {Heiselberg}, \citenamefont {Pethick}, \citenamefont {Smith},\ and\ \citenamefont {Viverit}}]{PhysRevLett.85.2418}%
  \BibitemOpen
  \bibfield  {author} {\bibinfo {author} {\bibfnamefont {H.}~\bibnamefont {Heiselberg}}, \bibinfo {author} {\bibfnamefont {C.~J.}\ \bibnamefont {Pethick}}, \bibinfo {author} {\bibfnamefont {H.}~\bibnamefont {Smith}}, \ and\ \bibinfo {author} {\bibfnamefont {L.}~\bibnamefont {Viverit}},\ }\href {\doibase 10.1103/PhysRevLett.85.2418} {\bibfield  {journal} {\bibinfo  {journal} {Phys. Rev. Lett.}\ }\textbf {\bibinfo {volume} {85}},\ \bibinfo {pages} {2418} (\bibinfo {year} {2000})}\BibitemShut {NoStop}%
\bibitem [{\citenamefont {Bijlsma}\ \emph {et~al.}(2000)\citenamefont {Bijlsma}, \citenamefont {Heringa},\ and\ \citenamefont {Stoof}}]{PhysRevA.61.053601}%
  \BibitemOpen
  \bibfield  {author} {\bibinfo {author} {\bibfnamefont {M.~J.}\ \bibnamefont {Bijlsma}}, \bibinfo {author} {\bibfnamefont {B.~A.}\ \bibnamefont {Heringa}}, \ and\ \bibinfo {author} {\bibfnamefont {H.~T.~C.}\ \bibnamefont {Stoof}},\ }\href {\doibase 10.1103/PhysRevA.61.053601} {\bibfield  {journal} {\bibinfo  {journal} {Phys. Rev. A}\ }\textbf {\bibinfo {volume} {61}},\ \bibinfo {pages} {053601} (\bibinfo {year} {2000})}\BibitemShut {NoStop}%
\bibitem [{\citenamefont {Klein}\ and\ \citenamefont {Fleischhauer}(2005)}]{PhysRevA.71.033605}%
  \BibitemOpen
  \bibfield  {author} {\bibinfo {author} {\bibfnamefont {A.}~\bibnamefont {Klein}}\ and\ \bibinfo {author} {\bibfnamefont {M.}~\bibnamefont {Fleischhauer}},\ }\href {\doibase 10.1103/PhysRevA.71.033605} {\bibfield  {journal} {\bibinfo  {journal} {Phys. Rev. A}\ }\textbf {\bibinfo {volume} {71}},\ \bibinfo {pages} {033605} (\bibinfo {year} {2005})}\BibitemShut {NoStop}%
\bibitem [{\citenamefont {Nakano}\ and\ \citenamefont {Yabu}(2016)}]{PhysRevB.93.205144}%
  \BibitemOpen
  \bibfield  {author} {\bibinfo {author} {\bibfnamefont {E.}~\bibnamefont {Nakano}}\ and\ \bibinfo {author} {\bibfnamefont {H.}~\bibnamefont {Yabu}},\ }\href {\doibase 10.1103/PhysRevB.93.205144} {\bibfield  {journal} {\bibinfo  {journal} {Phys. Rev. B}\ }\textbf {\bibinfo {volume} {93}},\ \bibinfo {pages} {205144} (\bibinfo {year} {2016})}\BibitemShut {NoStop}%
\bibitem [{\citenamefont {Camacho-Guardian}\ \emph {et~al.}(2018)\citenamefont {Camacho-Guardian}, \citenamefont {Pe\~na Ardila}, \citenamefont {Pohl},\ and\ \citenamefont {Bruun}}]{PhysRevLett.121.013401}%
  \BibitemOpen
  \bibfield  {author} {\bibinfo {author} {\bibfnamefont {A.}~\bibnamefont {Camacho-Guardian}}, \bibinfo {author} {\bibfnamefont {L.~A.}\ \bibnamefont {Pe\~na Ardila}}, \bibinfo {author} {\bibfnamefont {T.}~\bibnamefont {Pohl}}, \ and\ \bibinfo {author} {\bibfnamefont {G.~M.}\ \bibnamefont {Bruun}},\ }\href {\doibase 10.1103/PhysRevLett.121.013401} {\bibfield  {journal} {\bibinfo  {journal} {Phys. Rev. Lett.}\ }\textbf {\bibinfo {volume} {121}},\ \bibinfo {pages} {013401} (\bibinfo {year} {2018})}\BibitemShut {NoStop}%
\bibitem [{\citenamefont {Naidon}(2018)}]{naidon2018two}%
  \BibitemOpen
  \bibfield  {author} {\bibinfo {author} {\bibfnamefont {P.}~\bibnamefont {Naidon}},\ }\href@noop {} {\bibfield  {journal} {\bibinfo  {journal} {J. Phys. Soc. Jpn.}\ }\textbf {\bibinfo {volume} {87}},\ \bibinfo {pages} {043002} (\bibinfo {year} {2018})}\BibitemShut {NoStop}%
\bibitem [{\citenamefont {G{\ifmmode\acute{o}\else\'{o}\fi}mez-Lozada}\ \emph {et~al.}(2025)\citenamefont {G{\ifmmode\acute{o}\else\'{o}\fi}mez-Lozada}, \citenamefont {Hiyane}, \citenamefont {Busch},\ and\ \citenamefont {Fogarty}}]{Gomez-Lozada2025Apr}%
  \BibitemOpen
  \bibfield  {author} {\bibinfo {author} {\bibfnamefont {F.}~\bibnamefont {G{\ifmmode\acute{o}\else\'{o}\fi}mez-Lozada}}, \bibinfo {author} {\bibfnamefont {H.}~\bibnamefont {Hiyane}}, \bibinfo {author} {\bibfnamefont {T.}~\bibnamefont {Busch}}, \ and\ \bibinfo {author} {\bibfnamefont {T.}~\bibnamefont {Fogarty}},\ }\href {\doibase 10.1103/PhysRevResearch.7.023053} {\bibfield  {journal} {\bibinfo  {journal} {Phys. Rev. Res.}\ }\textbf {\bibinfo {volume} {7}},\ \bibinfo {pages} {023053} (\bibinfo {year} {2025})}\BibitemShut {NoStop}%
\bibitem [{\citenamefont {Hiyane}\ \emph {et~al.}(2024)\citenamefont {Hiyane}, \citenamefont {Busch},\ and\ \citenamefont {Fogarty}}]{Hiyane2024Aug}%
  \BibitemOpen
  \bibfield  {author} {\bibinfo {author} {\bibfnamefont {H.}~\bibnamefont {Hiyane}}, \bibinfo {author} {\bibfnamefont {T.}~\bibnamefont {Busch}}, \ and\ \bibinfo {author} {\bibfnamefont {T.}~\bibnamefont {Fogarty}},\ }\href {\doibase 10.1103/PhysRevResearch.6.L032040} {\bibfield  {journal} {\bibinfo  {journal} {Phys. Rev. Res.}\ }\textbf {\bibinfo {volume} {6}},\ \bibinfo {pages} {L032040} (\bibinfo {year} {2024})}\BibitemShut {NoStop}%
\bibitem [{\citenamefont {Hiyane}\ \emph {et~al.}(2025)\citenamefont {Hiyane}, \citenamefont {Fogarty}, \citenamefont {Pelayo},\ and\ \citenamefont {Busch}}]{hiyane2025condensate}%
  \BibitemOpen
  \bibfield  {author} {\bibinfo {author} {\bibfnamefont {H.}~\bibnamefont {Hiyane}}, \bibinfo {author} {\bibfnamefont {T.}~\bibnamefont {Fogarty}}, \bibinfo {author} {\bibfnamefont {J.~C.}\ \bibnamefont {Pelayo}}, \ and\ \bibinfo {author} {\bibfnamefont {T.}~\bibnamefont {Busch}},\ }\href@noop {} {\bibfield  {journal} {\bibinfo  {journal} {New J. Phys.}\ }\textbf {\bibinfo {volume} {27}},\ \bibinfo {pages} {124502} (\bibinfo {year} {2025})}\BibitemShut {NoStop}%
\bibitem [{\citenamefont {Ruderman}\ and\ \citenamefont {Kittel}(1954)}]{RudermanKittel1954}%
  \BibitemOpen
  \bibfield  {author} {\bibinfo {author} {\bibfnamefont {M.~A.}\ \bibnamefont {Ruderman}}\ and\ \bibinfo {author} {\bibfnamefont {C.}~\bibnamefont {Kittel}},\ }\href {\doibase 10.1103/PhysRev.96.99} {\bibfield  {journal} {\bibinfo  {journal} {Phys. Rev.}\ }\textbf {\bibinfo {volume} {96}},\ \bibinfo {pages} {99} (\bibinfo {year} {1954})}\BibitemShut {NoStop}%
\bibitem [{\citenamefont {Kasuya}(1956)}]{Kasuya1956}%
  \BibitemOpen
  \bibfield  {author} {\bibinfo {author} {\bibfnamefont {T.}~\bibnamefont {Kasuya}},\ }\href {\doibase 10.1143/PTP.16.45} {\bibfield  {journal} {\bibinfo  {journal} {Progress of Theoretical Physics}\ }\textbf {\bibinfo {volume} {16}},\ \bibinfo {pages} {45} (\bibinfo {year} {1956})}\BibitemShut {NoStop}%
\bibitem [{\citenamefont {Yosida}(1957)}]{Yosida1957}%
  \BibitemOpen
  \bibfield  {author} {\bibinfo {author} {\bibfnamefont {K.}~\bibnamefont {Yosida}},\ }\href {\doibase 10.1103/PhysRev.106.893} {\bibfield  {journal} {\bibinfo  {journal} {Physical Review}\ }\textbf {\bibinfo {volume} {106}},\ \bibinfo {pages} {893} (\bibinfo {year} {1957})}\BibitemShut {NoStop}%
\bibitem [{\citenamefont {Pethick}\ and\ \citenamefont {Smith}(2008)}]{pethick2008bose}%
  \BibitemOpen
  \bibfield  {author} {\bibinfo {author} {\bibfnamefont {C.~J.}\ \bibnamefont {Pethick}}\ and\ \bibinfo {author} {\bibfnamefont {H.}~\bibnamefont {Smith}},\ }\href@noop {} {\emph {\bibinfo {title} {Bose--Einstein condensation in dilute gases}}}\ (\bibinfo  {publisher} {Cambridge university press},\ \bibinfo {year} {2008})\BibitemShut {NoStop}%
\bibitem [{\citenamefont {Recati}\ \emph {et~al.}(2005)\citenamefont {Recati}, \citenamefont {Fuchs}, \citenamefont {Pe\ifmmode~\mbox{\c{c}}\else \c{c}\fi{}a},\ and\ \citenamefont {Zwerger}}]{PhysRevA.72.023616}%
  \BibitemOpen
  \bibfield  {author} {\bibinfo {author} {\bibfnamefont {A.}~\bibnamefont {Recati}}, \bibinfo {author} {\bibfnamefont {J.~N.}\ \bibnamefont {Fuchs}}, \bibinfo {author} {\bibfnamefont {C.~S.}\ \bibnamefont {Pe\ifmmode~\mbox{\c{c}}\else \c{c}\fi{}a}}, \ and\ \bibinfo {author} {\bibfnamefont {W.}~\bibnamefont {Zwerger}},\ }\href {\doibase 10.1103/PhysRevA.72.023616} {\bibfield  {journal} {\bibinfo  {journal} {Phys. Rev. A}\ }\textbf {\bibinfo {volume} {72}},\ \bibinfo {pages} {023616} (\bibinfo {year} {2005})}\BibitemShut {NoStop}%
\bibitem [{\citenamefont {Nishida}(2009)}]{PhysRevA.79.013629}%
  \BibitemOpen
  \bibfield  {author} {\bibinfo {author} {\bibfnamefont {Y.}~\bibnamefont {Nishida}},\ }\href {\doibase 10.1103/PhysRevA.79.013629} {\bibfield  {journal} {\bibinfo  {journal} {Phys. Rev. A}\ }\textbf {\bibinfo {volume} {79}},\ \bibinfo {pages} {013629} (\bibinfo {year} {2009})}\BibitemShut {NoStop}%
\bibitem [{\citenamefont {Huber}\ \emph {et~al.}(2019)\citenamefont {Huber}, \citenamefont {Hammer},\ and\ \citenamefont {Volosniev}}]{PhysRevResearch.1.033177}%
  \BibitemOpen
  \bibfield  {author} {\bibinfo {author} {\bibfnamefont {D.}~\bibnamefont {Huber}}, \bibinfo {author} {\bibfnamefont {H.-W.}\ \bibnamefont {Hammer}}, \ and\ \bibinfo {author} {\bibfnamefont {A.~G.}\ \bibnamefont {Volosniev}},\ }\href {\doibase 10.1103/PhysRevResearch.1.033177} {\bibfield  {journal} {\bibinfo  {journal} {Phys. Rev. Res.}\ }\textbf {\bibinfo {volume} {1}},\ \bibinfo {pages} {033177} (\bibinfo {year} {2019})}\BibitemShut {NoStop}%
\bibitem [{\citenamefont {Arg\"uello-Luengo}\ \emph {et~al.}(2022)\citenamefont {Arg\"uello-Luengo}, \citenamefont {Gonz\'alez-Tudela},\ and\ \citenamefont {Gonz\'alez-Cuadra}}]{PhysRevLett.129.083401}%
  \BibitemOpen
  \bibfield  {author} {\bibinfo {author} {\bibfnamefont {J.}~\bibnamefont {Arg\"uello-Luengo}}, \bibinfo {author} {\bibfnamefont {A.}~\bibnamefont {Gonz\'alez-Tudela}}, \ and\ \bibinfo {author} {\bibfnamefont {D.}~\bibnamefont {Gonz\'alez-Cuadra}},\ }\href {\doibase 10.1103/PhysRevLett.129.083401} {\bibfield  {journal} {\bibinfo  {journal} {Phys. Rev. Lett.}\ }\textbf {\bibinfo {volume} {129}},\ \bibinfo {pages} {083401} (\bibinfo {year} {2022})}\BibitemShut {NoStop}%
\bibitem [{\citenamefont {Akamatsu}\ \emph {et~al.}(2024)\citenamefont {Akamatsu}, \citenamefont {Endo}, \citenamefont {Fujii},\ and\ \citenamefont {Hongo}}]{PhysRevA.110.033304}%
  \BibitemOpen
  \bibfield  {author} {\bibinfo {author} {\bibfnamefont {Y.}~\bibnamefont {Akamatsu}}, \bibinfo {author} {\bibfnamefont {S.}~\bibnamefont {Endo}}, \bibinfo {author} {\bibfnamefont {K.}~\bibnamefont {Fujii}}, \ and\ \bibinfo {author} {\bibfnamefont {M.}~\bibnamefont {Hongo}},\ }\href {\doibase 10.1103/PhysRevA.110.033304} {\bibfield  {journal} {\bibinfo  {journal} {Phys. Rev. A}\ }\textbf {\bibinfo {volume} {110}},\ \bibinfo {pages} {033304} (\bibinfo {year} {2024})}\BibitemShut {NoStop}%
\bibitem [{\citenamefont {DeSalvo}\ \emph {et~al.}(2019)\citenamefont {DeSalvo}, \citenamefont {Patel}, \citenamefont {Cai},\ and\ \citenamefont {Chin}}]{desalvo2019observation}%
  \BibitemOpen
  \bibfield  {author} {\bibinfo {author} {\bibfnamefont {B.~J.}\ \bibnamefont {DeSalvo}}, \bibinfo {author} {\bibfnamefont {K.}~\bibnamefont {Patel}}, \bibinfo {author} {\bibfnamefont {G.}~\bibnamefont {Cai}}, \ and\ \bibinfo {author} {\bibfnamefont {C.}~\bibnamefont {Chin}},\ }\href@noop {} {\bibfield  {journal} {\bibinfo  {journal} {Nature}\ }\textbf {\bibinfo {volume} {568}},\ \bibinfo {pages} {61} (\bibinfo {year} {2019})}\BibitemShut {NoStop}%
\bibitem [{\citenamefont {Edri}\ \emph {et~al.}(2020)\citenamefont {Edri}, \citenamefont {Raz}, \citenamefont {Matzliah}, \citenamefont {Davidson},\ and\ \citenamefont {Ozeri}}]{PhysRevLett.124.163401}%
  \BibitemOpen
  \bibfield  {author} {\bibinfo {author} {\bibfnamefont {H.}~\bibnamefont {Edri}}, \bibinfo {author} {\bibfnamefont {B.}~\bibnamefont {Raz}}, \bibinfo {author} {\bibfnamefont {N.}~\bibnamefont {Matzliah}}, \bibinfo {author} {\bibfnamefont {N.}~\bibnamefont {Davidson}}, \ and\ \bibinfo {author} {\bibfnamefont {R.}~\bibnamefont {Ozeri}},\ }\href {\doibase 10.1103/PhysRevLett.124.163401} {\bibfield  {journal} {\bibinfo  {journal} {Phys. Rev. Lett.}\ }\textbf {\bibinfo {volume} {124}},\ \bibinfo {pages} {163401} (\bibinfo {year} {2020})}\BibitemShut {NoStop}%
\bibitem [{\citenamefont {Baroni}\ \emph {et~al.}(2024{\natexlab{b}})\citenamefont {Baroni}, \citenamefont {Huang}, \citenamefont {Fritsche}, \citenamefont {Dobler}, \citenamefont {Anich}, \citenamefont {Kirilov}, \citenamefont {Grimm}, \citenamefont {Bastarrachea-Magnani}, \citenamefont {Massignan},\ and\ \citenamefont {Bruun}}]{baroni2024mediated}%
  \BibitemOpen
  \bibfield  {author} {\bibinfo {author} {\bibfnamefont {C.}~\bibnamefont {Baroni}}, \bibinfo {author} {\bibfnamefont {B.}~\bibnamefont {Huang}}, \bibinfo {author} {\bibfnamefont {I.}~\bibnamefont {Fritsche}}, \bibinfo {author} {\bibfnamefont {E.}~\bibnamefont {Dobler}}, \bibinfo {author} {\bibfnamefont {G.}~\bibnamefont {Anich}}, \bibinfo {author} {\bibfnamefont {E.}~\bibnamefont {Kirilov}}, \bibinfo {author} {\bibfnamefont {R.}~\bibnamefont {Grimm}}, \bibinfo {author} {\bibfnamefont {M.~A.}\ \bibnamefont {Bastarrachea-Magnani}}, \bibinfo {author} {\bibfnamefont {P.}~\bibnamefont {Massignan}}, \ and\ \bibinfo {author} {\bibfnamefont {G.~M.}\ \bibnamefont {Bruun}},\ }\href@noop {} {\bibfield  {journal} {\bibinfo  {journal} {Nat. Phys.}\ }\textbf {\bibinfo {volume} {20}},\ \bibinfo {pages} {68} (\bibinfo {year} {2024}{\natexlab{b}})}\BibitemShut {NoStop}%
\bibitem [{\citenamefont {Cai}\ \emph {et~al.}(2026)\citenamefont {Cai}, \citenamefont {Ando}, \citenamefont {McCusker},\ and\ \citenamefont {Chin}}]{w7rd-2qpv}%
  \BibitemOpen
  \bibfield  {author} {\bibinfo {author} {\bibfnamefont {G.}~\bibnamefont {Cai}}, \bibinfo {author} {\bibfnamefont {H.}~\bibnamefont {Ando}}, \bibinfo {author} {\bibfnamefont {S.}~\bibnamefont {McCusker}}, \ and\ \bibinfo {author} {\bibfnamefont {C.}~\bibnamefont {Chin}},\ }\href {\doibase 10.1103/w7rd-2qpv} {\bibfield  {journal} {\bibinfo  {journal} {Phys. Rev. Lett.}\ }\textbf {\bibinfo {volume} {136}},\ \bibinfo {pages} {083403} (\bibinfo {year} {2026})}\BibitemShut {NoStop}%
\bibitem [{\citenamefont {Fr{\"o}hlich}(1954)}]{frohlich1954electrons}%
  \BibitemOpen
  \bibfield  {author} {\bibinfo {author} {\bibfnamefont {H.}~\bibnamefont {Fr{\"o}hlich}},\ }\href@noop {} {\bibfield  {journal} {\bibinfo  {journal} {Advances in Physics}\ }\textbf {\bibinfo {volume} {3}},\ \bibinfo {pages} {325} (\bibinfo {year} {1954})}\BibitemShut {NoStop}%
\bibitem [{\citenamefont {Bardeen}\ and\ \citenamefont {Pines}(1955)}]{PhysRev.99.1140}%
  \BibitemOpen
  \bibfield  {author} {\bibinfo {author} {\bibfnamefont {J.}~\bibnamefont {Bardeen}}\ and\ \bibinfo {author} {\bibfnamefont {D.}~\bibnamefont {Pines}},\ }\href {\doibase 10.1103/PhysRev.99.1140} {\bibfield  {journal} {\bibinfo  {journal} {Phys. Rev.}\ }\textbf {\bibinfo {volume} {99}},\ \bibinfo {pages} {1140} (\bibinfo {year} {1955})}\BibitemShut {NoStop}%
\bibitem [{\citenamefont {Devreese}\ and\ \citenamefont {Alexandrov}(2009)}]{devreese2009frohlich}%
  \BibitemOpen
  \bibfield  {author} {\bibinfo {author} {\bibfnamefont {J.~T.}\ \bibnamefont {Devreese}}\ and\ \bibinfo {author} {\bibfnamefont {A.~S.}\ \bibnamefont {Alexandrov}},\ }\href {\doibase 10.1088/0034-4885/72/6/066501} {\bibfield  {journal} {\bibinfo  {journal} {Rep. Prog. Phys.}\ }\textbf {\bibinfo {volume} {72}},\ \bibinfo {pages} {066501} (\bibinfo {year} {2009})}\BibitemShut {NoStop}%
\bibitem [{\citenamefont {Devreese}(2010)}]{Devreese2010Dec}%
  \BibitemOpen
  \bibfield  {author} {\bibinfo {author} {\bibfnamefont {J.~T.}\ \bibnamefont {Devreese}},\ }\href {\doibase 10.48550/arXiv.1012.4576} {\bibfield  {journal} {\bibinfo  {journal} {arXiv}\ } (\bibinfo {year} {2010}),\ 10.48550/arXiv.1012.4576},\ \Eprint {http://arxiv.org/abs/1012.4576} {1012.4576} \BibitemShut {NoStop}%
\bibitem [{\citenamefont {Filip}\ \emph {et~al.}(2021)\citenamefont {Filip}, \citenamefont {Haber},\ and\ \citenamefont {Neaton}}]{Filip2021Aug}%
  \BibitemOpen
  \bibfield  {author} {\bibinfo {author} {\bibfnamefont {M.~R.}\ \bibnamefont {Filip}}, \bibinfo {author} {\bibfnamefont {J.~B.}\ \bibnamefont {Haber}}, \ and\ \bibinfo {author} {\bibfnamefont {J.~B.}\ \bibnamefont {Neaton}},\ }\href {\doibase 10.1103/PhysRevLett.127.067401} {\bibfield  {journal} {\bibinfo  {journal} {Phys. Rev. Lett.}\ }\textbf {\bibinfo {volume} {127}},\ \bibinfo {pages} {067401} (\bibinfo {year} {2021})}\BibitemShut {NoStop}%
\bibitem [{\citenamefont {Tubman}\ \emph {et~al.}(2026)\citenamefont {Tubman}, \citenamefont {Coveney}, \citenamefont {Hsu}, \citenamefont {Montoya-Castillo}, \citenamefont {Filip}, \citenamefont {Neaton}, \citenamefont {Li}, \citenamefont {Vlcek},\ and\ \citenamefont {Alvertis}}]{Tubman2026Jun}%
  \BibitemOpen
  \bibfield  {author} {\bibinfo {author} {\bibfnamefont {N.~M.}\ \bibnamefont {Tubman}}, \bibinfo {author} {\bibfnamefont {C.~J.~N.}\ \bibnamefont {Coveney}}, \bibinfo {author} {\bibfnamefont {C.-E.}\ \bibnamefont {Hsu}}, \bibinfo {author} {\bibfnamefont {A.}~\bibnamefont {Montoya-Castillo}}, \bibinfo {author} {\bibfnamefont {M.~R.}\ \bibnamefont {Filip}}, \bibinfo {author} {\bibfnamefont {J.~B.}\ \bibnamefont {Neaton}}, \bibinfo {author} {\bibfnamefont {Z.}~\bibnamefont {Li}}, \bibinfo {author} {\bibfnamefont {V.}~\bibnamefont {Vlcek}}, \ and\ \bibinfo {author} {\bibfnamefont {A.~M.}\ \bibnamefont {Alvertis}},\ }\href {\doibase 10.1103/7hqv-hn2v} {\bibfield  {journal} {\bibinfo  {journal} {Phys. Rev. B}\ }\textbf {\bibinfo {volume} {113}},\ \bibinfo {pages} {245144} (\bibinfo {year} {2026})}\BibitemShut {NoStop}%
\bibitem [{\citenamefont {Watanabe}\ and\ \citenamefont {Murayama}(2012)}]{PhysRevLett.108.251602}%
  \BibitemOpen
  \bibfield  {author} {\bibinfo {author} {\bibfnamefont {H.}~\bibnamefont {Watanabe}}\ and\ \bibinfo {author} {\bibfnamefont {H.}~\bibnamefont {Murayama}},\ }\href {\doibase 10.1103/PhysRevLett.108.251602} {\bibfield  {journal} {\bibinfo  {journal} {Phys. Rev. Lett.}\ }\textbf {\bibinfo {volume} {108}},\ \bibinfo {pages} {251602} (\bibinfo {year} {2012})}\BibitemShut {NoStop}%
\bibitem [{\citenamefont {Hidaka}(2013)}]{PhysRevLett.110.091601}%
  \BibitemOpen
  \bibfield  {author} {\bibinfo {author} {\bibfnamefont {Y.}~\bibnamefont {Hidaka}},\ }\href {\doibase 10.1103/PhysRevLett.110.091601} {\bibfield  {journal} {\bibinfo  {journal} {Phys. Rev. Lett.}\ }\textbf {\bibinfo {volume} {110}},\ \bibinfo {pages} {091601} (\bibinfo {year} {2013})}\BibitemShut {NoStop}%
\bibitem [{\citenamefont {Duine}\ and\ \citenamefont {MacDonald}(2005)}]{PhysRevLett.95.230403}%
  \BibitemOpen
  \bibfield  {author} {\bibinfo {author} {\bibfnamefont {R.~A.}\ \bibnamefont {Duine}}\ and\ \bibinfo {author} {\bibfnamefont {A.~H.}\ \bibnamefont {MacDonald}},\ }\href {\doibase 10.1103/PhysRevLett.95.230403} {\bibfield  {journal} {\bibinfo  {journal} {Phys. Rev. Lett.}\ }\textbf {\bibinfo {volume} {95}},\ \bibinfo {pages} {230403} (\bibinfo {year} {2005})}\BibitemShut {NoStop}%
\bibitem [{\citenamefont {Sandri}\ \emph {et~al.}(2011)\citenamefont {Sandri}, \citenamefont {Minguzzi},\ and\ \citenamefont {Toigo}}]{Sandri2011}%
  \BibitemOpen
  \bibfield  {author} {\bibinfo {author} {\bibfnamefont {M.}~\bibnamefont {Sandri}}, \bibinfo {author} {\bibfnamefont {A.}~\bibnamefont {Minguzzi}}, \ and\ \bibinfo {author} {\bibfnamefont {F.}~\bibnamefont {Toigo}},\ }\href {\doibase 10.1209/0295-5075/96/66004} {\bibfield  {journal} {\bibinfo  {journal} {EPL}\ }\textbf {\bibinfo {volume} {96}},\ \bibinfo {pages} {66004} (\bibinfo {year} {2011})}\BibitemShut {NoStop}%
\bibitem [{\citenamefont {Tajima}\ and\ \citenamefont {Iida}(2021)}]{tajima2021non}%
  \BibitemOpen
  \bibfield  {author} {\bibinfo {author} {\bibfnamefont {H.}~\bibnamefont {Tajima}}\ and\ \bibinfo {author} {\bibfnamefont {K.}~\bibnamefont {Iida}},\ }\href@noop {} {\bibfield  {journal} {\bibinfo  {journal} {Journal of the Physical Society of Japan}\ }\textbf {\bibinfo {volume} {90}},\ \bibinfo {pages} {024004} (\bibinfo {year} {2021})}\BibitemShut {NoStop}%
\bibitem [{\citenamefont {Zhang}\ \emph {et~al.}(2023)\citenamefont {Zhang}, \citenamefont {Oue}, \citenamefont {Tajima}, \citenamefont {Matsuo},\ and\ \citenamefont {Liang}}]{PhysRevB.108.155303}%
  \BibitemOpen
  \bibfield  {author} {\bibinfo {author} {\bibfnamefont {T.}~\bibnamefont {Zhang}}, \bibinfo {author} {\bibfnamefont {D.}~\bibnamefont {Oue}}, \bibinfo {author} {\bibfnamefont {H.}~\bibnamefont {Tajima}}, \bibinfo {author} {\bibfnamefont {M.}~\bibnamefont {Matsuo}}, \ and\ \bibinfo {author} {\bibfnamefont {H.}~\bibnamefont {Liang}},\ }\href {\doibase 10.1103/PhysRevB.108.155303} {\bibfield  {journal} {\bibinfo  {journal} {Phys. Rev. B}\ }\textbf {\bibinfo {volume} {108}},\ \bibinfo {pages} {155303} (\bibinfo {year} {2023})}\BibitemShut {NoStop}%
\bibitem [{\citenamefont {Scazza}\ \emph {et~al.}(2017)\citenamefont {Scazza}, \citenamefont {Valtolina}, \citenamefont {Massignan}, \citenamefont {Recati}, \citenamefont {Amico}, \citenamefont {Burchianti}, \citenamefont {Fort}, \citenamefont {Inguscio}, \citenamefont {Zaccanti},\ and\ \citenamefont {Roati}}]{PhysRevLett.118.083602}%
  \BibitemOpen
  \bibfield  {author} {\bibinfo {author} {\bibfnamefont {F.}~\bibnamefont {Scazza}}, \bibinfo {author} {\bibfnamefont {G.}~\bibnamefont {Valtolina}}, \bibinfo {author} {\bibfnamefont {P.}~\bibnamefont {Massignan}}, \bibinfo {author} {\bibfnamefont {A.}~\bibnamefont {Recati}}, \bibinfo {author} {\bibfnamefont {A.}~\bibnamefont {Amico}}, \bibinfo {author} {\bibfnamefont {A.}~\bibnamefont {Burchianti}}, \bibinfo {author} {\bibfnamefont {C.}~\bibnamefont {Fort}}, \bibinfo {author} {\bibfnamefont {M.}~\bibnamefont {Inguscio}}, \bibinfo {author} {\bibfnamefont {M.}~\bibnamefont {Zaccanti}}, \ and\ \bibinfo {author} {\bibfnamefont {G.}~\bibnamefont {Roati}},\ }\href {\doibase 10.1103/PhysRevLett.118.083602} {\bibfield  {journal} {\bibinfo  {journal} {Phys. Rev. Lett.}\ }\textbf {\bibinfo {volume} {118}},\ \bibinfo {pages} {083602} (\bibinfo {year} {2017})}\BibitemShut {NoStop}%
\bibitem [{\citenamefont {Valtolina}\ \emph {et~al.}(2017)\citenamefont {Valtolina}, \citenamefont {Scazza}, \citenamefont {Amico}, \citenamefont {Burchianti}, \citenamefont {Recati}, \citenamefont {Enss}, \citenamefont {Inguscio}, \citenamefont {Zaccanti},\ and\ \citenamefont {Roati}}]{valtolina2017exploring}%
  \BibitemOpen
  \bibfield  {author} {\bibinfo {author} {\bibfnamefont {G.}~\bibnamefont {Valtolina}}, \bibinfo {author} {\bibfnamefont {F.}~\bibnamefont {Scazza}}, \bibinfo {author} {\bibfnamefont {A.}~\bibnamefont {Amico}}, \bibinfo {author} {\bibfnamefont {A.}~\bibnamefont {Burchianti}}, \bibinfo {author} {\bibfnamefont {A.}~\bibnamefont {Recati}}, \bibinfo {author} {\bibfnamefont {T.}~\bibnamefont {Enss}}, \bibinfo {author} {\bibfnamefont {M.}~\bibnamefont {Inguscio}}, \bibinfo {author} {\bibfnamefont {M.}~\bibnamefont {Zaccanti}}, \ and\ \bibinfo {author} {\bibfnamefont {G.}~\bibnamefont {Roati}},\ }\href@noop {} {\bibfield  {journal} {\bibinfo  {journal} {Nature Physics}\ }\textbf {\bibinfo {volume} {13}},\ \bibinfo {pages} {704} (\bibinfo {year} {2017})}\BibitemShut {NoStop}%
\bibitem [{\citenamefont {Scazza}\ \emph {et~al.}(2020)\citenamefont {Scazza}, \citenamefont {Valtolina}, \citenamefont {Amico}, \citenamefont {Tavares}, \citenamefont {Inguscio}, \citenamefont {Ketterle}, \citenamefont {Roati},\ and\ \citenamefont {Zaccanti}}]{PhysRevA.101.013603}%
  \BibitemOpen
  \bibfield  {author} {\bibinfo {author} {\bibfnamefont {F.}~\bibnamefont {Scazza}}, \bibinfo {author} {\bibfnamefont {G.}~\bibnamefont {Valtolina}}, \bibinfo {author} {\bibfnamefont {A.}~\bibnamefont {Amico}}, \bibinfo {author} {\bibfnamefont {P.~E.~S.}\ \bibnamefont {Tavares}}, \bibinfo {author} {\bibfnamefont {M.}~\bibnamefont {Inguscio}}, \bibinfo {author} {\bibfnamefont {W.}~\bibnamefont {Ketterle}}, \bibinfo {author} {\bibfnamefont {G.}~\bibnamefont {Roati}}, \ and\ \bibinfo {author} {\bibfnamefont {M.}~\bibnamefont {Zaccanti}},\ }\href {\doibase 10.1103/PhysRevA.101.013603} {\bibfield  {journal} {\bibinfo  {journal} {Phys. Rev. A}\ }\textbf {\bibinfo {volume} {101}},\ \bibinfo {pages} {013603} (\bibinfo {year} {2020})}\BibitemShut {NoStop}%
\bibitem [{\citenamefont {Ji}\ \emph {et~al.}(2022)\citenamefont {Ji}, \citenamefont {Schumacher}, \citenamefont {Assump\ifmmode \mbox{\c{c}}\else~\c{c}\fi{}\ ao}, \citenamefont {Chen}, \citenamefont {M\"akinen}, \citenamefont {Vivanco},\ and\ \citenamefont {Navon}}]{PhysRevLett.129.203402}%
  \BibitemOpen
  \bibfield  {author} {\bibinfo {author} {\bibfnamefont {Y.}~\bibnamefont {Ji}}, \bibinfo {author} {\bibfnamefont {G.~L.}\ \bibnamefont {Schumacher}}, \bibinfo {author} {\bibfnamefont {G.~G.~T.}\ \bibnamefont {Assump\ifmmode \mbox{\c{c}}\else~\c{c}\fi{}\ ao}}, \bibinfo {author} {\bibfnamefont {J.}~\bibnamefont {Chen}}, \bibinfo {author} {\bibfnamefont {J.~T.}\ \bibnamefont {M\"akinen}}, \bibinfo {author} {\bibfnamefont {F.~J.}\ \bibnamefont {Vivanco}}, \ and\ \bibinfo {author} {\bibfnamefont {N.}~\bibnamefont {Navon}},\ }\href {\doibase 10.1103/PhysRevLett.129.203402} {\bibfield  {journal} {\bibinfo  {journal} {Phys. Rev. Lett.}\ }\textbf {\bibinfo {volume} {129}},\ \bibinfo {pages} {203402} (\bibinfo {year} {2022})}\BibitemShut {NoStop}%
\bibitem [{\citenamefont {Recati}\ and\ \citenamefont {Stringari}(2022)}]{recati2022coherently}%
  \BibitemOpen
  \bibfield  {author} {\bibinfo {author} {\bibfnamefont {A.}~\bibnamefont {Recati}}\ and\ \bibinfo {author} {\bibfnamefont {S.}~\bibnamefont {Stringari}},\ }\href@noop {} {\bibfield  {journal} {\bibinfo  {journal} {Annu. Rev. Condens. Matter Phys.}\ }\textbf {\bibinfo {volume} {13}},\ \bibinfo {pages} {407} (\bibinfo {year} {2022})}\BibitemShut {NoStop}%
\bibitem [{\citenamefont {Yu}\ and\ \citenamefont {Yang}(2008)}]{PhysRevLett.100.090404}%
  \BibitemOpen
  \bibfield  {author} {\bibinfo {author} {\bibfnamefont {Y.}~\bibnamefont {Yu}}\ and\ \bibinfo {author} {\bibfnamefont {K.}~\bibnamefont {Yang}},\ }\href {\doibase 10.1103/PhysRevLett.100.090404} {\bibfield  {journal} {\bibinfo  {journal} {Phys. Rev. Lett.}\ }\textbf {\bibinfo {volume} {100}},\ \bibinfo {pages} {090404} (\bibinfo {year} {2008})}\BibitemShut {NoStop}%
\bibitem [{\citenamefont {Blaizot}\ \emph {et~al.}(2015)\citenamefont {Blaizot}, \citenamefont {Hidaka},\ and\ \citenamefont {Satow}}]{PhysRevA.92.063629}%
  \BibitemOpen
  \bibfield  {author} {\bibinfo {author} {\bibfnamefont {J.-P.}\ \bibnamefont {Blaizot}}, \bibinfo {author} {\bibfnamefont {Y.}~\bibnamefont {Hidaka}}, \ and\ \bibinfo {author} {\bibfnamefont {D.}~\bibnamefont {Satow}},\ }\href {\doibase 10.1103/PhysRevA.92.063629} {\bibfield  {journal} {\bibinfo  {journal} {Phys. Rev. A}\ }\textbf {\bibinfo {volume} {92}},\ \bibinfo {pages} {063629} (\bibinfo {year} {2015})}\BibitemShut {NoStop}%
\bibitem [{\citenamefont {Bradlyn}\ and\ \citenamefont {Gromov}(2016)}]{PhysRevA.93.033642}%
  \BibitemOpen
  \bibfield  {author} {\bibinfo {author} {\bibfnamefont {B.}~\bibnamefont {Bradlyn}}\ and\ \bibinfo {author} {\bibfnamefont {A.}~\bibnamefont {Gromov}},\ }\href {\doibase 10.1103/PhysRevA.93.033642} {\bibfield  {journal} {\bibinfo  {journal} {Phys. Rev. A}\ }\textbf {\bibinfo {volume} {93}},\ \bibinfo {pages} {033642} (\bibinfo {year} {2016})}\BibitemShut {NoStop}%
\bibitem [{\citenamefont {Blaizot}\ \emph {et~al.}(2017)\citenamefont {Blaizot}, \citenamefont {Hidaka},\ and\ \citenamefont {Satow}}]{PhysRevA.96.063617}%
  \BibitemOpen
  \bibfield  {author} {\bibinfo {author} {\bibfnamefont {J.-P.}\ \bibnamefont {Blaizot}}, \bibinfo {author} {\bibfnamefont {Y.}~\bibnamefont {Hidaka}}, \ and\ \bibinfo {author} {\bibfnamefont {D.}~\bibnamefont {Satow}},\ }\href {\doibase 10.1103/PhysRevA.96.063617} {\bibfield  {journal} {\bibinfo  {journal} {Phys. Rev. A}\ }\textbf {\bibinfo {volume} {96}},\ \bibinfo {pages} {063617} (\bibinfo {year} {2017})}\BibitemShut {NoStop}%
\bibitem [{\citenamefont {Tajima}\ \emph {et~al.}(2021)\citenamefont {Tajima}, \citenamefont {Hidaka},\ and\ \citenamefont {Satow}}]{TajimaHidakaSatow2021}%
  \BibitemOpen
  \bibfield  {author} {\bibinfo {author} {\bibfnamefont {H.}~\bibnamefont {Tajima}}, \bibinfo {author} {\bibfnamefont {Y.}~\bibnamefont {Hidaka}}, \ and\ \bibinfo {author} {\bibfnamefont {D.}~\bibnamefont {Satow}},\ }\href {\doibase 10.1103/PhysRevResearch.3.013035} {\bibfield  {journal} {\bibinfo  {journal} {Phys. Rev. Res.}\ }\textbf {\bibinfo {volume} {3}},\ \bibinfo {pages} {013035} (\bibinfo {year} {2021})}\BibitemShut {NoStop}%
\bibitem [{\citenamefont {Gazzillo}\ and\ \citenamefont {de~Melo}(2026)}]{gazzillo2026inverse}%
  \BibitemOpen
  \bibfield  {author} {\bibinfo {author} {\bibfnamefont {Z.}~\bibnamefont {Gazzillo}}\ and\ \bibinfo {author} {\bibfnamefont {C.~A.}\ \bibnamefont {de~Melo}},\ }\href@noop {} {\bibfield  {journal} {\bibinfo  {journal} {arXiv preprint arXiv:2606.07784}\ } (\bibinfo {year} {2026})}\BibitemShut {NoStop}%
\bibitem [{\citenamefont {Dzyaloshinsky}(1958)}]{dzyaloshinsky1958thermodynamic}%
  \BibitemOpen
  \bibfield  {author} {\bibinfo {author} {\bibfnamefont {I.}~\bibnamefont {Dzyaloshinsky}},\ }\href@noop {} {\bibfield  {journal} {\bibinfo  {journal} {J. Phys. Chem. Solids}\ }\textbf {\bibinfo {volume} {4}},\ \bibinfo {pages} {241} (\bibinfo {year} {1958})}\BibitemShut {NoStop}%
\bibitem [{\citenamefont {Moriya}(1960)}]{Moriya1960}%
  \BibitemOpen
  \bibfield  {author} {\bibinfo {author} {\bibfnamefont {T.}~\bibnamefont {Moriya}},\ }\href {\doibase 10.1103/PhysRev.120.91} {\bibfield  {journal} {\bibinfo  {journal} {Physical Review}\ }\textbf {\bibinfo {volume} {120}},\ \bibinfo {pages} {91} (\bibinfo {year} {1960})}\BibitemShut {NoStop}%
\bibitem [{\citenamefont {Monroe}\ \emph {et~al.}(2021)\citenamefont {Monroe}, \citenamefont {Campbell}, \citenamefont {Duan}, \citenamefont {Gong}, \citenamefont {Gorshkov}, \citenamefont {Hess}, \citenamefont {Islam}, \citenamefont {Kim}, \citenamefont {Linke}, \citenamefont {Pagano}, \citenamefont {Richerme}, \citenamefont {Senko},\ and\ \citenamefont {Yao}}]{RevModPhys.93.025001}%
  \BibitemOpen
  \bibfield  {author} {\bibinfo {author} {\bibfnamefont {C.}~\bibnamefont {Monroe}}, \bibinfo {author} {\bibfnamefont {W.~C.}\ \bibnamefont {Campbell}}, \bibinfo {author} {\bibfnamefont {L.-M.}\ \bibnamefont {Duan}}, \bibinfo {author} {\bibfnamefont {Z.-X.}\ \bibnamefont {Gong}}, \bibinfo {author} {\bibfnamefont {A.~V.}\ \bibnamefont {Gorshkov}}, \bibinfo {author} {\bibfnamefont {P.~W.}\ \bibnamefont {Hess}}, \bibinfo {author} {\bibfnamefont {R.}~\bibnamefont {Islam}}, \bibinfo {author} {\bibfnamefont {K.}~\bibnamefont {Kim}}, \bibinfo {author} {\bibfnamefont {N.~M.}\ \bibnamefont {Linke}}, \bibinfo {author} {\bibfnamefont {G.}~\bibnamefont {Pagano}}, \bibinfo {author} {\bibfnamefont {P.}~\bibnamefont {Richerme}}, \bibinfo {author} {\bibfnamefont {C.}~\bibnamefont {Senko}}, \ and\ \bibinfo {author} {\bibfnamefont {N.~Y.}\ \bibnamefont {Yao}},\ }\href {\doibase 10.1103/RevModPhys.93.025001} {\bibfield  {journal} {\bibinfo  {journal} {Rev. Mod. Phys.}\ }\textbf {\bibinfo {volume} {93}},\ \bibinfo {pages}
  {025001} (\bibinfo {year} {2021})}\BibitemShut {NoStop}%
\bibitem [{\citenamefont {Peter}\ \emph {et~al.}(2012)\citenamefont {Peter}, \citenamefont {M\"uller}, \citenamefont {Wessel},\ and\ \citenamefont {B\"uchler}}]{PhysRevLett.109.025303}%
  \BibitemOpen
  \bibfield  {author} {\bibinfo {author} {\bibfnamefont {D.}~\bibnamefont {Peter}}, \bibinfo {author} {\bibfnamefont {S.}~\bibnamefont {M\"uller}}, \bibinfo {author} {\bibfnamefont {S.}~\bibnamefont {Wessel}}, \ and\ \bibinfo {author} {\bibfnamefont {H.~P.}\ \bibnamefont {B\"uchler}},\ }\href {\doibase 10.1103/PhysRevLett.109.025303} {\bibfield  {journal} {\bibinfo  {journal} {Phys. Rev. Lett.}\ }\textbf {\bibinfo {volume} {109}},\ \bibinfo {pages} {025303} (\bibinfo {year} {2012})}\BibitemShut {NoStop}%
\bibitem [{\citenamefont {Chen}\ \emph {et~al.}(2025)\citenamefont {Chen}, \citenamefont {Emperauger}, \citenamefont {Bornet}, \citenamefont {Caleca}, \citenamefont {G{\ifmmode\acute{e}\else\'{e}\fi}ly}, \citenamefont {Bintz}, \citenamefont {Chatterjee}, \citenamefont {Liu}, \citenamefont {Barredo}, \citenamefont {Yao}, \citenamefont {Lahaye}, \citenamefont {Mezzacapo}, \citenamefont {Roscilde},\ and\ \citenamefont {Browaeys}}]{Chen2025Jun}%
  \BibitemOpen
  \bibfield  {author} {\bibinfo {author} {\bibfnamefont {C.}~\bibnamefont {Chen}}, \bibinfo {author} {\bibfnamefont {G.}~\bibnamefont {Emperauger}}, \bibinfo {author} {\bibfnamefont {G.}~\bibnamefont {Bornet}}, \bibinfo {author} {\bibfnamefont {F.}~\bibnamefont {Caleca}}, \bibinfo {author} {\bibfnamefont {B.}~\bibnamefont {G{\ifmmode\acute{e}\else\'{e}\fi}ly}}, \bibinfo {author} {\bibfnamefont {M.}~\bibnamefont {Bintz}}, \bibinfo {author} {\bibfnamefont {S.}~\bibnamefont {Chatterjee}}, \bibinfo {author} {\bibfnamefont {V.}~\bibnamefont {Liu}}, \bibinfo {author} {\bibfnamefont {D.}~\bibnamefont {Barredo}}, \bibinfo {author} {\bibfnamefont {N.~Y.}\ \bibnamefont {Yao}}, \bibinfo {author} {\bibfnamefont {T.}~\bibnamefont {Lahaye}}, \bibinfo {author} {\bibfnamefont {F.}~\bibnamefont {Mezzacapo}}, \bibinfo {author} {\bibfnamefont {T.}~\bibnamefont {Roscilde}}, \ and\ \bibinfo {author} {\bibfnamefont {A.}~\bibnamefont {Browaeys}},\ }\href {\doibase 10.1126/science.adn0618} {\bibfield  {journal} {\bibinfo  {journal}
  {Science}\ }\textbf {\bibinfo {volume} {389}},\ \bibinfo {pages} {483} (\bibinfo {year} {2025})}\BibitemShut {NoStop}%
\bibitem [{\citenamefont {Nishikawa}\ and\ \citenamefont {Saito}(2025)}]{hsbt-c46n}%
  \BibitemOpen
  \bibfield  {author} {\bibinfo {author} {\bibfnamefont {H.}~\bibnamefont {Nishikawa}}\ and\ \bibinfo {author} {\bibfnamefont {K.}~\bibnamefont {Saito}},\ }\href {\doibase 10.1103/hsbt-c46n} {\bibfield  {journal} {\bibinfo  {journal} {Phys. Rev. Lett.}\ }\textbf {\bibinfo {volume} {135}},\ \bibinfo {pages} {147102} (\bibinfo {year} {2025})}\BibitemShut {NoStop}%
\bibitem [{\citenamefont {Kawasaki}\ and\ \citenamefont {Danshita}(2026)}]{75k9-hx9k}%
  \BibitemOpen
  \bibfield  {author} {\bibinfo {author} {\bibfnamefont {D.}~\bibnamefont {Kawasaki}}\ and\ \bibinfo {author} {\bibfnamefont {I.}~\bibnamefont {Danshita}},\ }\href {\doibase 10.1103/75k9-hx9k} {\bibfield  {journal} {\bibinfo  {journal} {Phys. Rev. A}\ }\textbf {\bibinfo {volume} {113}},\ \bibinfo {pages} {053325} (\bibinfo {year} {2026})}\BibitemShut {NoStop}%
\bibitem [{sup()}]{suppmat}%
  \BibitemOpen
  \href@noop {} {}\bibinfo {note} {{See supplemental materials for further details }}\BibitemShut {NoStop}%
\bibitem [{\citenamefont {Scazza}\ \emph {et~al.}(2014)\citenamefont {Scazza}, \citenamefont {Hofrichter}, \citenamefont {H{\"o}fer}, \citenamefont {De~Groot}, \citenamefont {Bloch},\ and\ \citenamefont {F{\"o}lling}}]{scazza2014observation}%
  \BibitemOpen
  \bibfield  {author} {\bibinfo {author} {\bibfnamefont {F.}~\bibnamefont {Scazza}}, \bibinfo {author} {\bibfnamefont {C.}~\bibnamefont {Hofrichter}}, \bibinfo {author} {\bibfnamefont {M.}~\bibnamefont {H{\"o}fer}}, \bibinfo {author} {\bibfnamefont {P.~C.}\ \bibnamefont {De~Groot}}, \bibinfo {author} {\bibfnamefont {I.}~\bibnamefont {Bloch}}, \ and\ \bibinfo {author} {\bibfnamefont {S.}~\bibnamefont {F{\"o}lling}},\ }\href@noop {} {\bibfield  {journal} {\bibinfo  {journal} {Nature Physics}\ }\textbf {\bibinfo {volume} {10}},\ \bibinfo {pages} {779} (\bibinfo {year} {2014})}\BibitemShut {NoStop}%
\bibitem [{\citenamefont {Zhang}\ \emph {et~al.}(2016)\citenamefont {Zhang}, \citenamefont {Zhang}, \citenamefont {Cheng}, \citenamefont {Chen}, \citenamefont {Zhang},\ and\ \citenamefont {Zhai}}]{PhysRevA.93.043601}%
  \BibitemOpen
  \bibfield  {author} {\bibinfo {author} {\bibfnamefont {R.}~\bibnamefont {Zhang}}, \bibinfo {author} {\bibfnamefont {D.}~\bibnamefont {Zhang}}, \bibinfo {author} {\bibfnamefont {Y.}~\bibnamefont {Cheng}}, \bibinfo {author} {\bibfnamefont {W.}~\bibnamefont {Chen}}, \bibinfo {author} {\bibfnamefont {P.}~\bibnamefont {Zhang}}, \ and\ \bibinfo {author} {\bibfnamefont {H.}~\bibnamefont {Zhai}},\ }\href {\doibase 10.1103/PhysRevA.93.043601} {\bibfield  {journal} {\bibinfo  {journal} {Phys. Rev. A}\ }\textbf {\bibinfo {volume} {93}},\ \bibinfo {pages} {043601} (\bibinfo {year} {2016})}\BibitemShut {NoStop}%
\bibitem [{\citenamefont {Kan\'asz-Nagy}\ \emph {et~al.}(2018)\citenamefont {Kan\'asz-Nagy}, \citenamefont {Ashida}, \citenamefont {Shi}, \citenamefont {Moca}, \citenamefont {Ikeda}, \citenamefont {F\"olling}, \citenamefont {Cirac}, \citenamefont {Zar\'and},\ and\ \citenamefont {Demler}}]{PhysRevB.97.155156}%
  \BibitemOpen
  \bibfield  {author} {\bibinfo {author} {\bibfnamefont {M.}~\bibnamefont {Kan\'asz-Nagy}}, \bibinfo {author} {\bibfnamefont {Y.}~\bibnamefont {Ashida}}, \bibinfo {author} {\bibfnamefont {T.}~\bibnamefont {Shi}}, \bibinfo {author} {\bibfnamefont {C.~u. u. u. u. P. m.~c.}\ \bibnamefont {Moca}}, \bibinfo {author} {\bibfnamefont {T.~N.}\ \bibnamefont {Ikeda}}, \bibinfo {author} {\bibfnamefont {S.}~\bibnamefont {F\"olling}}, \bibinfo {author} {\bibfnamefont {J.~I.}\ \bibnamefont {Cirac}}, \bibinfo {author} {\bibfnamefont {G.}~\bibnamefont {Zar\'and}}, \ and\ \bibinfo {author} {\bibfnamefont {E.~A.}\ \bibnamefont {Demler}},\ }\href {\doibase 10.1103/PhysRevB.97.155156} {\bibfield  {journal} {\bibinfo  {journal} {Phys. Rev. B}\ }\textbf {\bibinfo {volume} {97}},\ \bibinfo {pages} {155156} (\bibinfo {year} {2018})}\BibitemShut {NoStop}%
\bibitem [{\citenamefont {Chen}\ \emph {et~al.}(2018)\citenamefont {Chen}, \citenamefont {Xu},\ and\ \citenamefont {You}}]{PhysRevA.98.023601}%
  \BibitemOpen
  \bibfield  {author} {\bibinfo {author} {\bibfnamefont {J.-J.}\ \bibnamefont {Chen}}, \bibinfo {author} {\bibfnamefont {Z.-F.}\ \bibnamefont {Xu}}, \ and\ \bibinfo {author} {\bibfnamefont {L.}~\bibnamefont {You}},\ }\href {\doibase 10.1103/PhysRevA.98.023601} {\bibfield  {journal} {\bibinfo  {journal} {Phys. Rev. A}\ }\textbf {\bibinfo {volume} {98}},\ \bibinfo {pages} {023601} (\bibinfo {year} {2018})}\BibitemShut {NoStop}%
\bibitem [{\citenamefont {Schmidt}\ \emph {et~al.}(2018)\citenamefont {Schmidt}, \citenamefont {Mayer}, \citenamefont {Bouton}, \citenamefont {Adam}, \citenamefont {Lausch}, \citenamefont {Spethmann},\ and\ \citenamefont {Widera}}]{PhysRevLett.121.130403}%
  \BibitemOpen
  \bibfield  {author} {\bibinfo {author} {\bibfnamefont {F.}~\bibnamefont {Schmidt}}, \bibinfo {author} {\bibfnamefont {D.}~\bibnamefont {Mayer}}, \bibinfo {author} {\bibfnamefont {Q.}~\bibnamefont {Bouton}}, \bibinfo {author} {\bibfnamefont {D.}~\bibnamefont {Adam}}, \bibinfo {author} {\bibfnamefont {T.}~\bibnamefont {Lausch}}, \bibinfo {author} {\bibfnamefont {N.}~\bibnamefont {Spethmann}}, \ and\ \bibinfo {author} {\bibfnamefont {A.}~\bibnamefont {Widera}},\ }\href {\doibase 10.1103/PhysRevLett.121.130403} {\bibfield  {journal} {\bibinfo  {journal} {Phys. Rev. Lett.}\ }\textbf {\bibinfo {volume} {121}},\ \bibinfo {pages} {130403} (\bibinfo {year} {2018})}\BibitemShut {NoStop}%
\bibitem [{\citenamefont {Tajima}(2025)}]{Tajima2025}%
  \BibitemOpen
  \bibfield  {author} {\bibinfo {author} {\bibfnamefont {H.}~\bibnamefont {Tajima}},\ }\href@noop {} {\emph {\bibinfo {title} {Quantum Field Theory in Condensed Matter: Feynman Diagrams from the Basics}}}\ (\bibinfo  {publisher} {Kyoritsu Shuppan},\ \bibinfo {address} {Tokyo},\ \bibinfo {year} {2025})\ \bibinfo {note} {in Japanese}\BibitemShut {NoStop}%
\bibitem [{\citenamefont {Tajima}\ \emph {et~al.}(2026)\citenamefont {Tajima}, \citenamefont {Nakano},\ and\ \citenamefont {Iida}}]{v1y3-2b6r}%
  \BibitemOpen
  \bibfield  {author} {\bibinfo {author} {\bibfnamefont {H.}~\bibnamefont {Tajima}}, \bibinfo {author} {\bibfnamefont {E.}~\bibnamefont {Nakano}}, \ and\ \bibinfo {author} {\bibfnamefont {K.}~\bibnamefont {Iida}},\ }\href {\doibase 10.1103/v1y3-2b6r} {\bibfield  {journal} {\bibinfo  {journal} {Phys. Rev. A}\ }\textbf {\bibinfo {volume} {113}},\ \bibinfo {pages} {L011305} (\bibinfo {year} {2026})}\BibitemShut {NoStop}%
\bibitem [{\citenamefont {Fetter}\ and\ \citenamefont {Walecka}(1971)}]{FetterWalecka1971}%
  \BibitemOpen
  \bibfield  {author} {\bibinfo {author} {\bibfnamefont {A.~L.}\ \bibnamefont {Fetter}}\ and\ \bibinfo {author} {\bibfnamefont {J.~D.}\ \bibnamefont {Walecka}},\ }\href@noop {} {\emph {\bibinfo {title} {Quantum Theory of Many-Particle Systems}}}\ (\bibinfo  {publisher} {McGraw-Hill},\ \bibinfo {address} {New York},\ \bibinfo {year} {1971})\BibitemShut {NoStop}%
\bibitem [{Note1()}]{Note1}%
  \BibitemOpen
  \bibinfo {note} {We note that $\Delta _j$ is the level splitting of the full localized single impurity Hamiltonian, including the bare energy difference between the two states, mean-field energy shifts from diagonal density-density coupling, and the one-body energy shift arising from $\ev *{\protect \hat H_{\protect \rm int}}_M$ induced by the exchange interaction.}\BibitemShut {Stop}%
\bibitem [{Note2()}]{Note2}%
  \BibitemOpen
  \bibinfo {note} {In the second line of Eq.~\protect \eqref {eq:j}, we use the analytic continuation from $\chi (\protect \bm {R},i\omega _n)=\DOTSI \intop \ilimits@ _0^{\beta }d\tau \protect \,e^{-i\omega _n\tau }\chi (\protect \bm {R},\tau )$ with the Matsubara frequency $\omega _n$ to the retarded susceptibility $\chi ^{\protect \rm ret}(\protect \bm {R},\omega )\equiv \chi (\protect \bm {R},i\omega _n\rightarrow \omega +i0^+)$ in the frequency representation.}\BibitemShut {Stop}%
\bibitem [{\citenamefont {Hayata}\ and\ \citenamefont {Hidaka}(2015)}]{PhysRevD.91.056006}%
  \BibitemOpen
  \bibfield  {author} {\bibinfo {author} {\bibfnamefont {T.}~\bibnamefont {Hayata}}\ and\ \bibinfo {author} {\bibfnamefont {Y.}~\bibnamefont {Hidaka}},\ }\href {\doibase 10.1103/PhysRevD.91.056006} {\bibfield  {journal} {\bibinfo  {journal} {Phys. Rev. D}\ }\textbf {\bibinfo {volume} {91}},\ \bibinfo {pages} {056006} (\bibinfo {year} {2015})}\BibitemShut {NoStop}%
\bibitem [{\citenamefont {Callaway}(1968)}]{PhysRev.170.576}%
  \BibitemOpen
  \bibfield  {author} {\bibinfo {author} {\bibfnamefont {J.}~\bibnamefont {Callaway}},\ }\href {\doibase 10.1103/PhysRev.170.576} {\bibfield  {journal} {\bibinfo  {journal} {Phys. Rev.}\ }\textbf {\bibinfo {volume} {170}},\ \bibinfo {pages} {576} (\bibinfo {year} {1968})}\BibitemShut {NoStop}%
\bibitem [{Note3()}]{Note3}%
  \BibitemOpen
  \bibinfo {note} {{ This condition is not strict and can be achieved, e.g.\ in ${}^{173}$Yb-${}^{174}$Yb or ${}^{40}$K-${}^{41}$K for the goldstino mode in the Bose--Fermi mixture~\cite {TajimaHidakaSatow2021} and for magnons in a Fermi--Fermi mixture~\cite {Sandri2011}, with inter-component repulsions.}}\BibitemShut {Stop}%
\bibitem [{\citenamefont {Zvonarev}\ \emph {et~al.}(2009)\citenamefont {Zvonarev}, \citenamefont {Cheianov},\ and\ \citenamefont {Giamarchi}}]{PhysRevLett.103.110401}%
  \BibitemOpen
  \bibfield  {author} {\bibinfo {author} {\bibfnamefont {M.~B.}\ \bibnamefont {Zvonarev}}, \bibinfo {author} {\bibfnamefont {V.~V.}\ \bibnamefont {Cheianov}}, \ and\ \bibinfo {author} {\bibfnamefont {T.}~\bibnamefont {Giamarchi}},\ }\href {\doibase 10.1103/PhysRevLett.103.110401} {\bibfield  {journal} {\bibinfo  {journal} {Phys. Rev. Lett.}\ }\textbf {\bibinfo {volume} {103}},\ \bibinfo {pages} {110401} (\bibinfo {year} {2009})}\BibitemShut {NoStop}%
\bibitem [{\citenamefont {Sekino}\ \emph {et~al.}(2024)\citenamefont {Sekino}, \citenamefont {Ominato}, \citenamefont {Tajima}, \citenamefont {Uchino},\ and\ \citenamefont {Matsuo}}]{PhysRevLett.133.163402}%
  \BibitemOpen
  \bibfield  {author} {\bibinfo {author} {\bibfnamefont {Y.}~\bibnamefont {Sekino}}, \bibinfo {author} {\bibfnamefont {Y.}~\bibnamefont {Ominato}}, \bibinfo {author} {\bibfnamefont {H.}~\bibnamefont {Tajima}}, \bibinfo {author} {\bibfnamefont {S.}~\bibnamefont {Uchino}}, \ and\ \bibinfo {author} {\bibfnamefont {M.}~\bibnamefont {Matsuo}},\ }\href {\doibase 10.1103/PhysRevLett.133.163402} {\bibfield  {journal} {\bibinfo  {journal} {Phys. Rev. Lett.}\ }\textbf {\bibinfo {volume} {133}},\ \bibinfo {pages} {163402} (\bibinfo {year} {2024})}\BibitemShut {NoStop}%
\bibitem [{\citenamefont {Sekino}\ \emph {et~al.}(2025)\citenamefont {Sekino}, \citenamefont {Ominato}, \citenamefont {Tajima}, \citenamefont {Uchino},\ and\ \citenamefont {Matsuo}}]{PhysRevA.111.033312}%
  \BibitemOpen
  \bibfield  {author} {\bibinfo {author} {\bibfnamefont {Y.}~\bibnamefont {Sekino}}, \bibinfo {author} {\bibfnamefont {Y.}~\bibnamefont {Ominato}}, \bibinfo {author} {\bibfnamefont {H.}~\bibnamefont {Tajima}}, \bibinfo {author} {\bibfnamefont {S.}~\bibnamefont {Uchino}}, \ and\ \bibinfo {author} {\bibfnamefont {M.}~\bibnamefont {Matsuo}},\ }\href {\doibase 10.1103/PhysRevA.111.033312} {\bibfield  {journal} {\bibinfo  {journal} {Phys. Rev. A}\ }\textbf {\bibinfo {volume} {111}},\ \bibinfo {pages} {033312} (\bibinfo {year} {2025})}\BibitemShut {NoStop}%
\bibitem [{\citenamefont {Miura}\ \emph {et~al.}(2024)\citenamefont {Miura}, \citenamefont {Shimomura},\ and\ \citenamefont {Totsuka}}]{PhysRevB.109.085141}%
  \BibitemOpen
  \bibfield  {author} {\bibinfo {author} {\bibfnamefont {U.}~\bibnamefont {Miura}}, \bibinfo {author} {\bibfnamefont {K.}~\bibnamefont {Shimomura}}, \ and\ \bibinfo {author} {\bibfnamefont {K.}~\bibnamefont {Totsuka}},\ }\href {\doibase 10.1103/PhysRevB.109.085141} {\bibfield  {journal} {\bibinfo  {journal} {Phys. Rev. B}\ }\textbf {\bibinfo {volume} {109}},\ \bibinfo {pages} {085141} (\bibinfo {year} {2024})}\BibitemShut {NoStop}%
\bibitem [{\citenamefont {Miura}\ and\ \citenamefont {Totsuka}(2025)}]{PhysRevB.111.075136}%
  \BibitemOpen
  \bibfield  {author} {\bibinfo {author} {\bibfnamefont {U.}~\bibnamefont {Miura}}\ and\ \bibinfo {author} {\bibfnamefont {K.}~\bibnamefont {Totsuka}},\ }\href {\doibase 10.1103/PhysRevB.111.075136} {\bibfield  {journal} {\bibinfo  {journal} {Phys. Rev. B}\ }\textbf {\bibinfo {volume} {111}},\ \bibinfo {pages} {075136} (\bibinfo {year} {2025})}\BibitemShut {NoStop}%
\bibitem [{\citenamefont {Reynolds}\ \emph {et~al.}(2020)\citenamefont {Reynolds}, \citenamefont {Schwartz}, \citenamefont {Ebling}, \citenamefont {Weyland}, \citenamefont {Brand},\ and\ \citenamefont {Andersen}}]{PhysRevLett.124.073401}%
  \BibitemOpen
  \bibfield  {author} {\bibinfo {author} {\bibfnamefont {L.~A.}\ \bibnamefont {Reynolds}}, \bibinfo {author} {\bibfnamefont {E.}~\bibnamefont {Schwartz}}, \bibinfo {author} {\bibfnamefont {U.}~\bibnamefont {Ebling}}, \bibinfo {author} {\bibfnamefont {M.}~\bibnamefont {Weyland}}, \bibinfo {author} {\bibfnamefont {J.}~\bibnamefont {Brand}}, \ and\ \bibinfo {author} {\bibfnamefont {M.~F.}\ \bibnamefont {Andersen}},\ }\href {\doibase 10.1103/PhysRevLett.124.073401} {\bibfield  {journal} {\bibinfo  {journal} {Phys. Rev. Lett.}\ }\textbf {\bibinfo {volume} {124}},\ \bibinfo {pages} {073401} (\bibinfo {year} {2020})}\BibitemShut {NoStop}%
\bibitem [{\citenamefont {Baym}\ and\ \citenamefont {Peng}(2025)}]{Baym2025Mar}%
  \BibitemOpen
  \bibfield  {author} {\bibinfo {author} {\bibfnamefont {G.}~\bibnamefont {Baym}}\ and\ \bibinfo {author} {\bibfnamefont {J.-C.}\ \bibnamefont {Peng}},\ }\href {\doibase 10.1016/j.nuclphysb.2025.116829} {\bibfield  {journal} {\bibinfo  {journal} {Nucl. Phys. B}\ }\textbf {\bibinfo {volume} {1012}},\ \bibinfo {pages} {116829} (\bibinfo {year} {2025})}\BibitemShut {NoStop}%
\end{thebibliography}%

\end{document}


\begin{CJK}{UTF8}{min}

\title{Supplemental Material: Tunable Mediated Interactions Near Spontaneous Symmetry Breaking: \\
From
Yukawa to Coulomb and Dzyaloshinskii--Moriya Interactions
}

\author{Hoshu Hiyane}
\affiliation{Institut f\"{u}r Theoretische Physik, Leibniz Universit\"{a}t Hannover, Appelstr. 2, 30167 Hannover, Germany}

\author{Hiroyuki Tajima}
\affiliation{Department of Physics, The University of Tokyo, Tokyo 113-0033, Japan}
\affiliation{RIKEN Nishina Center, Wako 351-0198, Japan}
\affiliation{Quark Nuclear Science Institute, The University of Tokyo, Tokyo 113-0033, Japan}

\maketitle
\end{CJK}

\section{Impurity Hamiltonian}

For completeness, we show the general Hamiltonian of the two 
impurities:
\begin{align}
    \label{eq:ItenerantImp_Ham}
    \hat H_{\rm I}
    &=\sum_{\sigma=\uparrow,\downarrow}
    \int d\bm{r}\,\hat{\psi}_{{\rm I},\sigma}^\dag(\bm{r})\left(-\frac{\nabla^2}{2m_\sigma}-\mu_\sigma\right)
    \hat{\psi}_{{\rm I},\sigma}(\bm{r})
    +\sum_{\sigma,\sigma'}g^{\sigma,\sigma'}_{\rm II}\int d\bm{r}\,
    \hat{\psi}_{{\rm I},\sigma}^\dag(\bm{r})
        \hat{\psi}_{{\rm I},\sigma'}^\dag(\bm{r})
            \hat{\psi}_{{\rm I},\sigma'}(\bm{r})
                \hat{\psi}_{{\rm I},\sigma}(\bm{r}),\\
                \notag
    &\quad
    +\sum_{\sigma,\sigma'}g^{\sigma,\sigma'}_{\rm IM}
    \int d\bm{r}\,
    \hat\rho_{{\rm I},\sigma}
    \hat\rho_{{\rm M},\sigma'},
\end{align}
where 
$\hat\rho_{{\rm I(M)},\sigma}=\hat\psi^\dag_{{\rm I(M)},\sigma}\hat\psi_{{\rm I(M)},\sigma}$ is the density operator for impurity (bath) particles.
For two bosonic impurities, the $s$-wave interaction with a strength $g^{\sigma\sigma'}_{\rm II}$ are allowed for any $\sigma=\{\uparrow,\downarrow\}$, while for two fermionic impurities, $g^{\uparrow\uparrow}_{\rm II}=g^{\downarrow\downarrow}_{\rm II}=0$ due to the Pauli blocking.
We note that $g_{\rm II}$ and the impurity-medium density-density coupling $g_{\rm IM}^{\sigma,\sigma'}$ may affect the diagonal part of $H_{\rm eff}$ 
the off-diagonal part is insensitive to them.
Thus, our result for probing $J(\bm{R})$ is robust against the density-density couplings.

In the main text, we discuss three kinds of systems altogether (see Tab.~1). 
Here, we list the explicit form of the Hamiltonian of the Fermi-Fermi and Bose-Fermi mixtures and their Nambu--Goldstone (NG) modes considered in this work.

\section{Magnon in a repulsive Fermi--Fermi mixture}
First, we discuss the Fermi--Fermi mixture case.
The Bose--Bose mixture also follows the same kind of Hamiltonians introduced below with bosonic symmetry. 
For a repulsive two-component Fermi gas with pseudospin $\sigma=\{\uparrow,\downarrow\}$, the relevant medium operator is the transverse spin density,
\begin{align}
\label{eq:ladder_operator_FermiMedia}
\hat S_{{\rm M},-}(\bm r)=\hat\psi_{\rm M\downarrow}^\dagger(\bm r)\hat\psi_{\rm M\uparrow}(\bm r),\quad
\hat S_{{\rm M},+}(\bm r)=\hat\psi_{\rm M\uparrow}^\dagger(\bm r)\hat\psi_{\rm M\downarrow}(\bm r).
\end{align}
The interaction with localized impurity spins is given by
\begin{align}
\hat{H}_{\rm int}=g\sum_{i=1,2}
\hat{S}_{{\rm I},+}(\bm{R}_i)\hat{S}_{{\rm M},-}(\bm R_i)+{\rm h.c.}
\label{eq:kondo}
\end{align}
The induced effective interaction reads
\begin{align}
\hat{V}_{\rm eff}=J(\bm{R})
\hat{S}_{{\rm I},+}(\bm{R}_1)\hat{S}_{{\rm I},-}(\bm{R}_2)+{\rm h.c.}
\end{align}
Close to a ferromagnetic instability, the transverse spin response is enhanced, and a magnon pole gives a large and long-ranged spin-flip interaction~\cite{Sandri2011}.

For the itinerant ferromagnetic case,
The medium Hamiltonian is given by
\begin{align}
    \label{eq:MediumHam_FFMix}
    \hat{H}_{\rm M}&=\sum_{\sigma=\uparrow,\downarrow}
    \int d\bm{r}\,\hat{\psi}_{{\rm M},\sigma}^\dag(\bm{r})\left(-\frac{\nabla^2}{2m}-\mu_\sigma\right)
    \hat{\psi}_{{\rm M},\sigma}(\bm{r})
    +U\int d\bm{r}\,
    \hat{\psi}_{{\rm M},\uparrow}^\dag(\bm{r})
        \hat{\psi}_{{\rm M},\downarrow}^\dag(\bm{r})
            \hat{\psi}_{{\rm M},\downarrow}(\bm{r})
                \hat{\psi}_{{\rm M},\uparrow}(\bm{r}),
\end{align}
where $U>0$ is the repulsive interaction strength, and we assume that a mass $m$ is spin-independent.
$\mu_\sigma$ is the spin-dependent chemical potential.

Using the random phase approximation (RPA)~\cite{PhysRev.170.576,Sandri2011,tajima2021non,PhysRevB.108.155303} with $\mu_\uparrow\geq \mu_\downarrow$, one can obtain the low-energy parameter of a magnon in Eq.~(13) in the main text 
as
\begin{align}
    Z=&\rho_\uparrow-\rho_{\downarrow},\\
    \omega_{\rm gap}=&\mu_{\uparrow}-\mu_{\downarrow},\\
    D=&\frac{1}{2m(\rho_\uparrow-\rho_\downarrow)}
    \left[\rho-\frac{\rho_\uparrow k_{\rm F,\uparrow }-\rho_\downarrow k_{\rm F,\downarrow} }{5mU(\rho_\uparrow-\rho_\downarrow)}\right],
\end{align}
where $\rho_{\sigma}$ and $k_{\rm F,\sigma}=(6\pi^2\rho_\sigma)^{1/3}$ are the spin-dependent number density and the Fermi momentum, respectively. 
Here, the total medium density is given by $\rho=\rho_{\uparrow}+\rho_{\downarrow}$.

\section{Goldstino in a repulsive Bose--Fermi mixture}
The system of a nearly supersymmetric Bose--Fermi mixture can be thought of similarly to the Fermi--Fermi mixture discussed above.
The medium Hamiltonian here is in the form of
\begin{align}
    \hat{H}_{\rm M}
    &=\int d\bm{r}
    \hat{\psi}_{{\rm M},b}^\dag(\bm{r})
    \left(-\frac{\nabla^2}{2m_b}-\mu_b\right)\hat{\psi}_{{\rm M},b}(\bm{r})
    ,
    +\int d\bm{r}
    \hat{\psi}_{{\rm M},f}^\dag(\bm{r})
    \left(-\frac{\nabla^2}{2m_f}-\mu_f\right)\hat{\psi}_{{\rm M},f}(\bm{r})\cr
    &\,+U_{bf}\int d\bm{r}\,
    \hat{\psi}_{{\rm M},b}^\dag(\bm{r})
        \hat{\psi}_{{\rm M},f}^\dag(\bm{r})
            \hat{\psi}_{{\rm M},f}(\bm{r})
                \hat{\psi}_{{\rm M},b}(\bm{r})
                +\frac{U_{bb}}{2}\int d\bm{r}\,
    \hat{\psi}_{{\rm M},b}^\dag(\bm{r})
        \hat{\psi}_{{\rm M},b}^\dag(\bm{r})
            \hat{\psi}_{{\rm M},b}(\bm{r})
                \hat{\psi}_{{\rm M},b}(\bm{r}),
\end{align}
where $\mu_{b(f)}$ is the bosonic (fermionic) chemical potential, $m_{b(f)}$ is the bosonic (fermionic) mass, $U_{bf(bb)}$ is the boson-fermion (boson-boson) coupling, and $\hat{\psi}_{{\rm M},b(f)}(\bm{r})$ is a bosonic (fermionic) operator in a Bose--Fermi mixture.

For Bose--Fermi mixture, the exchange interaction considered in Eq.~(3) main text can explicitly be written as
\begin{align}
    \hat H_{\rm int}=g\int d\bm r
    \hat \psi^\dagger_{\rm I,b}(\bm r)
    \hat \psi^\dagger_{\rm M,f}(\bm r)
    \hat \psi_{\rm I,f}(\bm r)
    \hat \psi_{\rm M,b}(\bm r)
    +{\rm h.c.}
\end{align}
This interaction Hamiltonian can be written in terms of the supercharge operators defined by
\begin{align}
    \hat{Q}_{\rm M}^\dag(\bm{r})
    =&\hat{\psi}_{{\rm M},f}^\dag(\bm{r}) \hat{\psi}_{{\rm M},b}(\bm{r}),\\
    \hat{Q}_{\rm M}(\bm{r})
    =&\hat{\psi}_{{\rm M},b}^\dag(\bm{r}) \hat{\psi}_{{\rm M},f}(\bm{r}).
\end{align}
The operator $\hat{Q}^{(\dagger)}_{\rm M}(\bm{r})$ flip the species from bosonic (fermionic) to fermionic (bosonic) and formally equivalent to the ladder operator $\hat{S}^{(\dagger)}_{\rm M,-}(\bm{r})$ introduced in the fermionic systems (see Eq.~\eqref{eq:ladder_operator_FermiMedia} also Eq.~(3) in the main text).

For localized impurities, the exchange process between Fermi and Bose atoms is described by
\begin{align}
\hat H_{\rm int}=g\sum_{i=1,2}
\hat Q_{\rm I}^\dag(\bm{R_i})\hat Q_{\rm M}(\bm R_i)+{\rm h.c.},
\label{eq:goldstinocoupling}
\end{align}
where $\hat Q_{\rm I}(\bm{r})=\hat{\psi}_{{\rm I},b}^\dag(\bm{r})\hat{\psi}_{{\rm I},f}(\bm{r})$ converts a fermionic impurity to a bosonic one. 
Then, the mediated two-impurity interaction is given by
\begin{align}
\hat{V}_{\rm eff}=J^*(\bm{R})\hat{Q}_{{\rm I}}^\dag(\bm{R}_1)
\hat{Q}_{{\rm I}}(\bm{R}_2)+{\rm h.c.}
\end{align}
Instead of the spin-flip susceptibility in a repulsive Fermi-Fermi mixture,
$J(\bm{R})$ is proportional to the goldstino spectrum. 
For a small mass imbalance ($m_f\simeq m_b$),
the RPA spectrum is characterized by $Z$, $\omega_{\rm gap}$, and $D$ in Eq.~(13) 
as~\cite{PhysRevLett.100.090404,PhysRevA.92.063629,PhysRevA.93.033642,PhysRevA.96.063617,TajimaHidakaSatow2021,gazzillo2026inverse}
\begin{align}
    Z=&\rho_b+\rho_f,\\
    \omega_{\rm gap}=&\mu_b-\mu_f+2\rho_b(U_{bf}-U_{bb}),\\
    D=&\frac{1}{2m_f},
\end{align}
where $\rho_{b(f)}$ is the bosonic (fermionic) number density.

\section{Dzyaloshinskii--Moriya-type interactions}
Here we show that $\hat{V}_{\rm eff}$ contains the Dzyaloshinskii--Moriya (DM)-type interactions when $J(\bm{R})$ is complex. 
Using
    $\hat{s}^{\pm}=\hat{s}^{x}\pm i\hat{s}^{y}$,
we obtain
\begin{align}
    \hat{s}_{1}^{-}\hat{s}_{2}^{+}
    &=
    (\hat{s}_{1}^{x}-i\hat{s}_{1}^{y})
    (\hat{s}_{2}^{x}+i\hat{s}_{2}^{y})
    \nonumber\\
    &=
    \hat{s}_{1}^{x}\hat{s}_{2}^{x}
    +\hat{s}_{1}^{y}\hat{s}_{2}^{y}
    +i\left(
    \hat{s}_{1}^{x}\hat{s}_{2}^{y}
    -\hat{s}_{1}^{y}\hat{s}_{2}^{x}
    \right),
\end{align}
Similarly, we obtain 
\begin{align}
    \hat{s}_{1}^{+}\hat{s}_{2}^{-}
    &=
    (\hat{s}_{1}^{x}+i\hat{s}_{1}^{y})
    (\hat{s}_{2}^{x}-i\hat{s}_{2}^{y})
    \nonumber\\
    &=
    \hat{s}_{1}^{x}\hat{s}_{2}^{x}
    +\hat{s}_{1}^{y}\hat{s}_{2}^{y}
    -i\left(
    \hat{s}_{1}^{x}\hat{s}_{2}^{y}
    -\hat{s}_{1}^{y}\hat{s}_{2}^{x}
    \right).
\end{align}
Writing
    $J(\bm{R})
    =
    {\rm Re}\,J(\bm{R})
    +i\,{\rm Im}\,J(\bm{R})$,
the effective interaction
    $\hat{V}_{\rm eff}
    \simeq
    J(\bm{R})\hat{s}_{1}^{-}\hat{s}_{2}^{+}
    +J^{*}(\bm{R})\hat{s}_{1}^{+}\hat{s}_{2}^{-}$
can be rewritten as
\begin{align}
    \hat{V}_{\rm eff}
    \simeq\;&
    \left[
    {\rm Re}\,J(\bm{R})
    +i\,{\rm Im}\,J(\bm{R})
    \right]
    \left[
    \hat{s}_{1}^{x}\hat{s}_{2}^{x}
    +\hat{s}_{1}^{y}\hat{s}_{2}^{y}
    +i\left(
    \hat{s}_{1}^{x}\hat{s}_{2}^{y}
    -\hat{s}_{1}^{y}\hat{s}_{2}^{x}
    \right)
    \right]
    +
    \left[
    {\rm Re}\,J(\bm{R})
    -i\,{\rm Im}\,J(\bm{R})
    \right]
    \left[
    \hat{s}_{1}^{x}\hat{s}_{2}^{x}
    +\hat{s}_{1}^{y}\hat{s}_{2}^{y}
    -i\left(
    \hat{s}_{1}^{x}\hat{s}_{2}^{y}
    -\hat{s}_{1}^{y}\hat{s}_{2}^{x}
    \right)
    \right]
    \nonumber\\
    =\;&
    2{\rm Re}\,J(\bm{R})
    \left(
    \hat{s}_{1}^{x}\hat{s}_{2}^{x}
    +\hat{s}_{1}^{y}\hat{s}_{2}^{y}
    \right)
    -
    2{\rm Im}\,J(\bm{R})
    \left(
    \hat{s}_{1}^{x}\hat{s}_{2}^{y}
    -\hat{s}_{1}^{y}\hat{s}_{2}^{x}
    \right).
\end{align}
Since
    $\left(
    \hat{\bm{s}}_{1}\times\hat{\bm{s}}_{2}
    \right)_{z}
    =
    \hat{s}_{1}^{x}\hat{s}_{2}^{y}
    -\hat{s}_{1}^{y}\hat{s}_{2}^{x}$,
we finally obtain
\begin{align}
    \hat{V}_{\rm eff}
    \simeq\;&
    2{\rm Re}\,J(\bm{R})
    \left(
    \hat{s}_{1}^{x}\hat{s}_{2}^{x}
    +\hat{s}_{1}^{y}\hat{s}_{2}^{y}
    \right)
    \nonumber\\
    &-
    2{\rm Im}\,J(\bm{R})
    \left(
    \hat{\bm{s}}_{1}\times\hat{\bm{s}}_{2}
    \right)_{z}.
\end{align}
Thus, $2{\rm Re}\,J(\bm{R})$ gives the $XY$ exchange coupling,
whereas $-2{\rm Im}\,J(\bm{R})$ takes the form of a
DM-type coupling.

\bibliographystyle{apsrev4-1}
\bibliography{Manuscript}